\documentclass[11pt]{article}
\usepackage{jheppub}
\usepackage[utf8]{inputenc}
\usepackage{amsfonts, amsmath, amsthm, amssymb}     
\usepackage{bbold}
\usepackage{mathtools, cuted}      
\usepackage{commath}
\usepackage[mathscr]{euscript}
\usepackage{color}
\usepackage{physics}
\usepackage{tensor}
\usepackage{hyperref}
\usepackage{multirow}
\usepackage[dvipsnames]{xcolor}
\usepackage{bclogo}
\usepackage{comment}
\usepackage{dsfont}
\usepackage{cancel}
\usepackage[normalem]{ulem} 
\usepackage{nicefrac}
\usepackage{appendix}
\usepackage{microtype}
\usepackage{etoolbox}
\usepackage{orcidlink}

        \hypersetup{
           breaklinks=true,   
           colorlinks=true,   
        }

\usepackage{tikz}
    \usetikzlibrary{decorations.pathreplacing}
    \usetikzlibrary{decorations.pathmorphing} 
    \usetikzlibrary{decorations.markings} 
    \usetikzlibrary{arrows.meta,bending}
    \usetikzlibrary{decorations}
    \usetikzlibrary{shapes.geometric}
    \usetikzlibrary{shapes}
    \usetikzlibrary{positioning}

\numberwithin{equation}{section}

\newcommand\numberthis{\addtocounter{equation}{1}\tag{\theequation}} 

\newcommand{\Op}{\mathcal{O}}

\newcommand{\h}{\mathbb{h}}
\newcommand{\p}{\partial}
\newcommand{\Disc}{\mathbb{D}}

\newcommand{\BSpace}{\mathcal{H}L^2_{\h}(\Disc)}

\renewcommand{\d}{\operatorname{d}\!}

\newcommand{\hbb}[0]{\mathbb{h}}
\newcommand{\Deltabb}[0]{\mathbb{\Delta}}

\begin{document}

\title{\vspace*{2cm} Conformal Blocks in Two and Four Dimensions from Oscillator Representations}

\author{Martin Ammon\,\orcidlink{0000-0002-3838-5968},}
\author{Jakob Hollweck\,\orcidlink{0000-0001-8454-2568},}
\author{Tobias Hössel\,\orcidlink{0009-0006-0582-1176},}
\author{Katharina W\"olfl\,\orcidlink{0009-0000-4380-2722}}
\affiliation{Theoretisch-Physikalisches Institut, Friedrich-Schiller-Universit\"at Jena,\\
Max-Wien-Platz 1, D-07743 Jena, Germany}

\emailAdd{martin.ammon@uni-jena.de}
\emailAdd{jakob.hollweck@uni-jena.de}
\emailAdd{tobias.hoessel@uni-jena.de}
\emailAdd{katharina.woelfl@uni-jena.de}

\vspace{1cm}

\abstract{The explicit computation of higher-point conformal blocks in any dimension is usually a challenging task. For two-dimensional conformal field theories in Euclidean signature, the oscillator formalism proves to be very efficient. We demonstrate this by reproducing the general $n$-point global conformal block in the comb channel in an elegant and direct manner. Exploiting similarities to the representation theory of two-dimensional CFTs, we extend the oscillator formalism to the computation of higher-point conformal blocks in four Euclidean dimensions. As a proof of concept, we explicitly compute the scalar four-point block with scalar exchange within this framework and discuss the extension to the higher-point case.}

\setcounter{tocdepth}{2} 
\maketitle


\addtocontents{toc}{\protect\setcounter{tocdepth}{1}}
\section{Introduction}
Conformal symmetry imposes strong constraints on correlation functions in general dimensions. For example, they are uniquely determined up to three points for scalar fields, apart from the CFT data. 
Starting at four points, this is no longer the case; however, one still can decompose such a higher-point correlation function into conformal blocks using the operator product expansion. Demanding consistency for this decomposition for four-point correlation functions even leads to constraints on the CFT data, a fact that eventually gave rise to the conformal bootstrap program \cite{Ferrara:1973yt,Polyakov:1974gs,Rattazzi:2008pe,El-Showk:2012cjh}. 

Currently, this program is being explored in the context of higher-point correlation functions. Even though crossing symmetry equations for higher-point correlation functions do not present new consistency conditions, one can view crossing symmetry of an $n$-point correlation function as infinitely many four-point consistency conditions \cite{Antunes:2020yzv,Buric:2021yak}. This is part of the reason why the study of higher-point conformal blocks is very topical
\cite{Simmons-Duffin:2012juh, Rosenhaus:2018zqn,Parikh:2019ygo, Fortin:2019dnq, Goncalves:2019znr, Parikh:2019dvm, Fortin:2019zkm,Irges:2020lgp,Pal:2020dqf, Hoback:2020pgj,Fortin:2020zxw,Poland:2021xjs,Cho:2017oxl, Fortin:2020yjz,Fortin:2020bfq,Haehl2020,Buric:2020dyz,Buric:2021ywo, Buric:2021ttm, Buric:2021kgy, Fortin:2023xqq}. 

The study of conformal blocks is closely connected to the one of conformal partial waves (CPWs), which are characterised as the result of a harmonic decomposition of a higher-point correlation function with respect to the conformal symmetry, see e.\,g.\ \cite{Dobrev:1977qv}. They can be computed through the shadow(-operator) formalism, first introduced in a series of publications \cite{Ferrara1972_0,Ferrara1972_1,Ferrara1972_2,Ferrara1972_3}, where one introduces for every operator $\mathcal{O}_{\Delta}(x)$ in the theory a shadow operator $\widetilde{\mathcal{O}}_{\widetilde{\Delta}}(x)$. It is  intrinsically non-local and has dimension $\widetilde{\Delta} = d-\Delta$ in a $d$-dimensional Euclidean CFT. The CPWs are eventually calculated by first inserting a shadow projector, which involves both $\mathcal{O}_{\Delta}(x)$ and $\widetilde{\mathcal{O}}_{\widetilde{\Delta}}(x)$, and then evaluating the integral. Consequently, such a conformal partial wave involves both a conformal block and its shadow block. Imposing fall-off conditions allows to extract the conformal block. More recent important work in the context of (higher-point) conformal blocks using the shadow formalism is for example \cite{Simmons-Duffin:2012juh, Rosenhaus:2018zqn}. 

Alternatively, it is possible to derive general properties of conformal partial waves from studying the Casimir equations, as was done in \cite{Dolan:2000ut,Dolan:2003hv,Dolan:2011dv}. These are constructed simply as eigenvalue equations of the Casimir  operators of the conformal algebra. For the case of $d>2$ and $n>4$ points, however, it turns out that the number of commuting Casimir  operators is strictly smaller than the number of cross-ratios, such that they do not characterise the associated CPW (or conformal blocks) completely anymore. This problem has been recently solved in the series of papers \cite{Buric:2020dyz,Buric:2021ywo, Buric:2021ttm, Buric:2021kgy}. There, it was shown how to construct a set of commuting Casimir operators including novel additional operators that account for the choice of tensor structure at the vertices through a limit of the Gaudin integrable model. The result holds for higher-point conformal blocks in higher dimensions for any channel.  

The approach we are presenting here, however, is closer in spirit to the shadow formalism. The original idea relies on the fact that the highest-weight representations of the Virasoro algebra can be formulated in terms of differential operators on a function space with basis functions given by monomials of so-called oscillator variables \cite{Gervais:1984hw, Zam1986}. This has recently been reviewed in \cite{Besken:2019jyw,Kraus2020}. At least for the highest-weight representations of its global subalgebra, this function space is a holomorphic function space over the complex open unit disk $\mathbb{D}$, a weighted Bergman space. Such a space can for example be constructed by the means of generalised coherent states \cite{Perelomov1986}.

In this work, we extend the oscillator method to the analysis of higher-point conformal blocks and, in particular, the $n$-point comb block. A major advantage of the oscillator formalism lies in its toolbox nature: we establish a small set of ingredients, called oscillator wavefunctions, from which higher-point conformal blocks can be computed in a constructive and straightforward manner. In contrast to the Casimir equations of a generic conformal block, which can be very hard to solve, the defining differential equations of these wavefunctions are very feasible and the solution can usually be found by an educated guess. Moreover, the method allows for an intuitive diagrammatic interpretation.

A second aim of this work is to generalise this oscillator construction to four Euclidean dimensions. To this end, we discuss highest-weight representations of $\text{SU}(2,2)$, the (four-cover of the) Lorentzian conformal group in four dimensions. Whereas the integrals that come with the insertion of a shadow projector can generally be very challenging to evaluate (even numerically \cite{Buric:2021yak}), we will find that here the involved integrals are by construction over products of orthogonal functions because of the employed generalised coherent states. An additional advantage is that we directly determine the conformal block and do not need to subtract the shadow block contribution. This is because physical positivity is implemented by construction, in contrast to the shadow integral where to ensure completeness of the conformal partial waves, one necessarily has to include contributions from other representations such as the principal series, which has complex weights \cite{Gadde:2017sjg}. 

The generalised coherent states of $\text{SU}(2,2)$ we are using have already been discussed by various authors. Initially introduced by Graev \cite{Graev1954} in 1954, the first applications in the context of field theory were by Rühl \cite{Ruehl:1972jy,Ruehl1973,Rühl1974}; they have also played a role in for example \cite{Haba:1975pk,Perelomov1986,Calixto_2011,Calixto:2010pm,Calixto:2014sfa}, which includes more recent work. 

This paper is structured as follows: in section \ref{sec:construction_oscillator_representation} we construct oscillator representations for Verma modules of the global conformal algebra in both two and four dimensions. We derive and subsequently solve the defining differential equations of the oscillator wavefunctions and introduce a diagrammatic language in section \ref{sec:derivation_solution_oscillator_eqs}. 
Using these wavefunctions, we compute the two- and three-point correlation functions in section \ref{sec:two_three_point_correlation_fct}. 
Section \ref{sec:conf_blocks_comb_channel} begins with the calculation of the $n$-point comb channel conformal block in two dimensions and the scalar four-point block with scalar exchange in four dimensions. We then discuss first a special case of the five-point and then $n$-point comb block that allows for resummation to showcase applicability for higher-point blocks. The section concludes with a summary and an outlook on various interesting future directions. 

In order to compare the constructions in two and four dimensions, each section introduces first the two-dimensional case and in the following subsection the four-dimensional one, highlighting similarities and differences, as the latter approach is more involved.

\section{Construction of Oscillator Representations}
\label{sec:construction_oscillator_representation}
We construct unitary irreducible representations for the Verma module of the (global) conformal algebra in both two and four dimensions by means of coherent states. These \emph{oscillator representations} are defined on a weighted Bergman space over homogeneous spaces of $\text{SU}(1,1)$ and $\text{SU}(2,2)$, respectively. 
The existence of a reproducing kernel and its expansion in terms of an orthogonal basis is a key feature for later explicit calculations, e.g.\ of conformal blocks.

\subsection{Oscillator Representations in Two Dimensions}
The Lorentzian global conformal group in two dimension is given by $\text{SO}(2,2)$, its Lie algebra splits into two sectors, each being isomorphic to $\mathfrak{su}(1,1)$. It is customary to call them holomorphic and anti-holomorphic sector and focus on only one of them. Since our aim is to compute conformal blocks in Euclidean space, we are in the following considering the corresponding highest-weight representations of the Euclidean conformal algebra.\footnote{These representations of the Euclidean conformal algebra are not unitary but only correspond to unitary representations of the Lorentzian conformal algebra after Wick rotation, which is why they are sometimes dubbed ``physical representations'' \cite{Gadde:2017sjg}.}

The Verma module of each of those sectors can be represented as the subspace of holomorphic functions of $L^2(\mathbb{D})$, i.e.\ as weighted Bergman spaces $\mathcal{H}L^2_\hbb(\mathbb{D})$ that we label by the conformal weight $\hbb$.
This is naturally a unitary irreducible representation of the $\mathfrak{su}(1,1)$ discrete series due to the isomorphism of the complex unit disc $\mathbb{D} \cong \text{SU}(1,1)/\text{U}(1)$. 

To be more explicit, the weighted Bergman spaces are Hilbert spaces with respect to the inner product\footnote{The constant prefactor is chosen in such a way that the constant 1-function has unit norm. Note that we chose the measure $\d^2 u = 2\d x\d y$ where $u=x+iy$ and $\d x$ as well as $\d y$ is the usual Lebesgue measure on the real line.}
\begin{equation}
    (f,g) = \frac{2\mathbb{h}-1}{2\pi}\int_\mathbb{D}\frac{\d^2 u}{(1-u\bar{u})^{2-2\mathbb{h}}}\, \overline{f(u)}g(u) \equiv \int_\mathbb{D} [\d u]_{\h}\, \overline{f(u)}g(u)\label{eq:SL2R_inner_product}
\end{equation}
that explicitly depends on the conformal weight $\h$. Our notation is based on \cite{Kraus2020} and \cite{Hall2000}. The weighted-$L^2$ condition precisely requires finite induced norms $\norm{f}<\infty$. Indeed, for $\h>\nicefrac{1}{2}$, the monomials $\phi_m(u) = u^m$ form an orthogonal basis since the inner product gives 
\begin{equation}
    (u^m,u^n) = \frac{n!}{(2\mathbb{h})_n}\,\delta_{m,n}\,. \label{eq:2d_orthogonality_relation}
\end{equation}
Here $(2\mathbb{h})_n = \frac{\Gamma(2\hbb+n)}{\Gamma(2\hbb)}$ is the Pochhammer symbol. 

A well-know feature arising from the fact that point-evaluations $f\mapsto f(u)$ are linear functionals on the Bergman space is the unique reproducing kernel $K_{\h}(u^\prime,\bar{u})$. Its defining property is the reproducing identity 
\begin{equation}
    f(u^\prime) = \int_\mathbb{D}[\d u]_\hbb \,K_{\h}(u^\prime,\bar{u}) f(u) \label{eq:2d_reproducing_identity}
\end{equation}
from which we obtain an integral representation of the projector. For the weighted Bergman space $\mathcal{H}L^2_{\h}(\Disc)$, this reproducing kernel takes the compact form
\begin{equation}
    K_{\h}(u^\prime,\bar{u}) = (1-u^\prime\bar{u})^{-2\h}\,.
\end{equation}

An oscillator representation of one sector of the global conformal algebra, namely $\mathfrak{sl}(2,\mathbb{R})$ which is isomorphic to $\mathfrak{su}(1,1)$, maps its generators $L_n$ to differential operators $\mathfrak{L}_n$ for $n=0,\pm 1$, 
\begin{align}
\label{eq:osc_gen_2d}
    \mathfrak{L}_1 &= \p_u\,, & \mathfrak{L}_0 &= u\p_u + \h\,, & \mathfrak{L}_{-1} &= u^2\p_u + 2\h u\,,
\end{align}
acting on functions of the Bergman space. This defines a highest-weight representation with highest-weight state $\phi_{\h}(u) = 1$ that is an eigenfunction of $\mathfrak{L}_0$ to the eigenvalue $\h$ and is annihilated by $\mathfrak{L}_1$. Iteratively acting with the lowering operator $\mathfrak{L}_{-1}$ gives monomials which form an orthogonal basis of the representation space. One can directly check the adjoint relation $(\mathfrak{L}_n)^\dagger = \mathfrak{L}_{-n}$ that holds with respect to the inner product of the Bergman space. 

To highlight the Verma module structure of the oscillator representation, let us use the familiar bra-ket notation: an $\mathfrak{sl}(2,\mathbb{R})$-Verma module is spanned by a highest-weight state $\ket{\h}$ and its descendants $\ket{\h,n} = (L_{-1})^n\ket{\h}$. We can draw the connection to the holomorphic functions spanning the Bergman space with the help of generalised coherent states  
\begin{equation}
     \ket{\bar{u}} \equiv \ket{\bar{u}}_{\hbb} = \mathrm{e}^{\bar{u}L_{-1}} \ket{\h} = \sum_{n=0}^\infty \frac{\bar{u}^n}{n!}\ket{\hbb,n}\,.
\end{equation} 
Clearly, the wavefunctions $\braket{u}{\hbb,n}$ reproduce the monomial basis of the Bergman space up to normalisation.
Note that the  states $\ket{\bar{u}}$ are chosen to be anti-holomorphic in the oscillator variable $u$ such that the wavefunctions $\braket{u}{f}=f(u)$ are holomorphic for any given vector $\ket{f}$ in the Verma module. This choice is in agreement with the inner product 
\begin{equation}
    \braket{f}{g} = \mel{f}{\mathbb{P}_{\h}}{g} = \int_\mathbb{D} [\d u]_\hbb \braket{ f}{\Bar{u}} \braket{u}{g} = \int_\mathbb{D} [\d u]_\hbb\, \overline{f(u)}g(u) = (f,g)\,,
\end{equation}
where we used $\mathbb{1} = \mathbb{P}_{\h} = \int [\d u]_\hbb \dyad{u}{\bar{u}}$ when restricting from the whole CFT Hilbert space to a specific Bergman space $\mathcal{H}L^2_{\h}(\Disc)$.
We encode the action of the generators on the states via $\mel{u}{L_n}{f} = \mathfrak{L}_n\braket{u}{f} = \mathfrak{L}_n f(u)$ in terms of the oscillator representation on holomorphic functions. 

\subsection{Oscillator Representations in Four Dimensions}\label{ssec:4d_Oscillator}
In four dimensions, the Lorentzian conformal group is given by $\text{SO}(4,2)$. Its Lie algebra is isomorphic to $\mathfrak{su}(2,2)$ and we construct, analogously to two dimensions, a unitary re\-pre\-sen\-ta\-tion by considering the quotient space of $\text{SU}(2,2)$ by its maximally compact subgroup\footnote{Note that not only the Lie algebras are the same, but also that $\mathbb{D}_4$ coincides with $\text{SO}(4,2)/(\text{SO}(4) \times \text{SO}(2))$ as discussed in \cite{coquereaux1990}.}
\begin{equation}
\label{eq:su22_quotient}
    \mathbb{D}_4 \cong \text{SU}(2,2)/(\text{SU}(2) \times \text{SU}(2) \times \text{U}(1))\,.
\end{equation}
This is a natural matrix-valued generalisation of the two dimensional unit disk $\mathbb{D}$ to $(2\times 2)$-matrices $U$ such that $\mathbb{1} - U U^\dagger>0$, i.\,e.\ $\mathbb{1} - U U^\dagger$ is positive-definite. Maybe more intuitively, $\mathbb{D}_4$ is an open ball of eight real dimensions (with respect to the Lie norm).  For more details on $\mathbb{D}_4$, see for example \cite{Calixto_2011,coquereaux1990}. 

One sees directly that a representation of $\mathfrak{su}(2,2)$, which is constructed by induction of the maximally compact subgroup in equation \eqref{eq:su22_quotient}, is characterised by the scaling dimension $\Deltabb$ and the two spins $j_1$ and $j_2$. However, we restrict ourselves here to the scalar case $j_1 = j_2 = 0$ such that we consider $\mathcal{H}L^2_{\Deltabb}(\Disc_4)$. Its inner product is given by
\begin{align}
    (f,g) &= c_\Deltabb \int_{\Disc_4} \frac{\d U}{\mathrm{det}^{4-\Deltabb}(\mathbb{1}-UU^\dagger)} \overline{f(U)}g(U) \equiv \int_{\mathbb{D}_4}[\d U]_\Deltabb\, \overline{f(U)}g(U) \label{eq:D4_inner_product} 
\end{align}
where $\d U$ denotes the Lebesgue measure on $\mathbb{C}^{2\times 2}$ and the prefactor 
\begin{equation}
    c_{\mathbb{\Delta}} = \frac{1}{\pi^4}(\mathbb{\Delta}-1)(\mathbb{\Delta}-2)^2(\mathbb{\Delta}-3)
\end{equation}
is chosen such that the constant 1-function has unit norm. Although this definition seems problematic for e.\,g. $\Deltabb=3$, it can be extended to all values $\Deltabb \geq 2$ \cite{Ruehl:1972jy}. We base our notation on \cite{Calixto_2011,Calixto:2014sfa}.

In the following, we work in a basis of orthogonal homogeneous polynomials for the weighted Bergman space. The basis functions 
\begin{equation}
    \phi_{q_a,q_b}^{j,m} (U) = \text{det}^m(U) \mathcal{D}^j_{q_a,q_b}(U)
\end{equation}
can be seen as enhancement of the Wigner $\mathcal{D}$-matrices for $\text{SU}(2)$ (holomorphically extended to $\text{SL}(2,\mathbb{C})$)
\begin{equation}
    \mathcal{D}^j_{q_a,q_b}(U) = \sqrt{\frac{(j + q_a)!(j-q_a)!}{(j+q_b)!(j - q_b)!}} \sum\limits_k \binom{j + q_b}{k} \binom{j-q_b}{k - q_a - q_b}
    u^k_{11} u^{j+q_a - k}_{12} u^{j+q_b - k}_{21} u^{k-q_a - q_b}_{22}\,,
\end{equation}
where $k$ runs from $\max(0,q_a + q_b)$ to $\min(j+q_a,j+q_b)$. The parameters $m\in\mathbb{N}_{0}$ and $j\in \mathbb{N}_{0}/2$ are independent, while $q_a,q_b=-j,-j+1,\dots,j$. As was shown in \cite{Calixto_2011}, the basis functions obey the orthogonality relation
\begin{align}
    ( \phi_{q_a,q_b}^{j,m} , \phi_{q_a',q_b'}^{j',m'}) = \int_{\mathbb{D}_4}[\d U]_\mathbb{\Delta} \overline{\phi_{q_a,q_b}^{j,m} (U)}\, \phi_{q_a',q_b'}^{j',m'} (U) = (\mathcal{N}^{j,m}_\mathbb{\Delta})^{-2}\delta_{j,j'}\delta_{m,m'}\delta_{q_a,q_a'}\delta_{q_b,q_b'}\label{eq:4d_orthonormality_property} 
    \end{align}
    where the normalisation is given by
    \begin{align} \mathcal{N}^{j,m}_\mathbb{\Delta} = \sqrt{\frac{2j+1}{\mathbb{\Delta} -1} \binom{m+\mathbb{\Delta} -2}{\mathbb{\Delta} - 2} \binom{m+2j+\mathbb{\Delta} - 1}{\mathbb{\Delta} - 2}}\,.
\end{align}
Similarly to the construction in two dimensions, this Bergman space gives rise to a reproducing kernel $K_\mathbb{\Delta}(U,U')$ that can be expanded in terms of the basis functions
\begin{align}
    K_\mathbb{\Delta}(U,U') &= \mathrm{det}^{-\mathbb{\Delta}}(\mathbb{1}-U^\dagger U')\notag\\ 
    &= \sum\limits_{j \in \mathbb{N}/2}\sum\limits_{m=0}^\infty \sum\limits_{q_a,q_b=-j}^j (\mathcal{N}^{j,m}_\mathbb{\Delta})^2 \overline{\phi_{q_a,q_b}^{j,m} (U)} \phi_{q_a,q_b}^{j,m} (U') \label{eq:expansion_kernel}\\
    &\equiv\sum\limits_{q_a,q_b}^{j,m} (\mathcal{N}^{j,m}_\mathbb{\Delta})^2 \overline{\phi_{q_a,q_b}^{j,m} (U)} \phi_{q_a,q_b}^{j,m} (U') \notag
\end{align}
which is defined for $\Deltabb \in \mathbb{N}$ and $\Deltabb \geq 2$ and where we introduced short-hand notation for the sum in the last step.

To define the oscillator representation in four dimensions, we first have to represent the matrix domain $\Disc_4$ in coordinates using the Pauli matrices $\sigma^\mu$ as basis for $(2\times 2)$-matrices. In particular, we choose coordinates $w_\mu$ to parameterise the points of the Bergman domain as
\begin{equation}
    U=w_0\sigma^0 - iw_1\sigma^1 - iw_2\sigma^2- iw_3\sigma^3\,.  \label{eq:def_U_matrix}
\end{equation}
This allows us to represent the generators $D,P^\mu,K^\mu$  and $M^{\mu\nu}$ of the conformal algebra as differential operators acting on $\mathcal{H}L^2_{\Deltabb}(\Disc_4)$
\begin{align}
\begin{split}\label{eq:4d_oscillator-gen}
    \mathfrak{P}^\mu = \partial^\mu\,, \quad \mathfrak{K}^\mu = w^2 \partial^\mu - 2w^\mu w_\rho \partial^\rho - 2w^\mu \mathbb{\Delta}\,,  \\
    \mathfrak{D} = w_\mu \partial^\mu + \mathbb{\Delta} \,, \qquad \mathfrak{M}^{\mu \nu} = w^\mu \partial^\nu - w^\nu \partial^\mu
    \,,
\end{split}
\end{align}
with $w^2 = w_\nu w^\nu$. Note that the generators obey the commutation relations 
\begin{subequations}
\label{eq:comm_relations_euclidPlus}
\begin{alignat}{3}
    &[\mathfrak{P}^\mu,\,\mathfrak{P}^\nu]=0\,,\qquad &&[\mathfrak{D},\,\mathfrak{P}^\mu]=-\mathfrak{P}^\mu\,,\qquad &&[\mathfrak{M}^{\mu\nu},\,\mathfrak{P}^\rho]=-\delta^{\mu\rho}\mathfrak{P}^\nu+\delta^{\nu\rho}\mathfrak{P}^\mu\,,\\
    &[\mathfrak{K}^\mu,\,\mathfrak{K}^\nu]=0\,,\quad &&[\mathfrak{D},\,\mathfrak{K}^\mu]=\mathfrak{K}^\mu\,, \quad &&[\mathfrak{M}^{\mu\nu},\,\mathfrak{K}^\rho]=-\delta^{\mu\rho}\mathfrak{K}^\nu+\delta^{\nu\rho}\mathfrak{K}^\mu\,,\\
    &[\mathfrak{M}^{\mu\nu},\,\mathfrak{D}]=0\,,&&[\mathfrak{K}^\mu,\,\mathfrak{P}^\nu]=2(\mathfrak{M}^{\mu\nu}+\delta^{\mu\nu}\mathfrak{D})\,,&&\\
    &[\mathfrak{M}^{\mu\nu},\mathfrak{M}^{\rho\sigma}]=-\delta^{\mu\rho}&&\mathfrak{M}^{\nu\sigma}-\delta^{\nu\sigma}\mathfrak{M}^{\mu\rho}+\delta^{\nu\rho}\mathfrak{M}^{\mu\sigma}+&&\,\delta^{\mu\sigma}\mathfrak{M}^{\nu\rho}\,.
\end{alignat}
\end{subequations}
Moreover, the generators fulfil the adjoint relations 
\begin{align}
\label{eq:adjoint_relations_4d}
    \left(\mathfrak{P}^\mu\right)^\dagger&=-\mathfrak{K}^\mu &
    \left(\mathfrak{M}^{\mu\nu}\right)^\dagger&=-\mathfrak{M}^{\mu\nu} &
    \mathfrak{D}^\dagger&=\mathfrak{D}
\end{align}
with respect to the inner product \eqref{eq:D4_inner_product}. This can be explicitly checked using the action of the generators on the basis functions, see appendix \ref{app:action_generators_basis}. 
Note that equations \eqref{eq:comm_relations_euclidPlus} allow for Euclidean highest-weight representations that, after Wick rotation, correspond to unitary irreducible highest-weight representations of $\mathfrak{su}(2,2)$.

Quite similarly to the oscillator representation in two dimensions, our analogue construction in four dimensions has the structure of a (generalised) Verma module. Typically, one defines the higher-dimensional Verma module $\mathcal{V}_\Deltabb$ starting from a highest-weight state $\ket{\Deltabb}$ that is annihilated by all generators $K^\mu$ of special conformal transformations (SCT). Its descendants, arising from iterative action of translation generators $P^\mu$, span the module. However, in our representation the constant highest-weight wavefunction $\phi^{0,0}_{0,0} = \braket{U}{\Deltabb} =1$ is annihilated by all derivatives, i.\,e. $\mathfrak{P}_\mu 1 = 0$, and the SCT generators span the representation space
\begin{equation}
    \mathcal{H}L^2_\Deltabb(\mathbb{D}_4) = \text{span}\left\{\prod \nolimits_{\mu = 0}^3 \mathfrak{K}^{n_\mu}_\mu 1: n_\mu = 0,1,\dots\right\}\,.
\end{equation}
Hence, the connection between the ``abstract'' generator $L\in\{D,P^\mu,M^{\mu\nu},K^\mu\}$ acting on vectors and the corresponding differential operator $\mathfrak{L}\in\{\mathfrak{D},\mathfrak{P}^\mu,\mathfrak{M}^{\mu\nu},\mathfrak{K}^\mu\}$ acting on wavefunctions must read
\begin{equation}
    \label{eq:draw_out_generator_4d}
    \mel{U}{L}{f} = \mathfrak{L}^\dagger \braket{U}{f} = \mathfrak{L}^\dagger f(U)\,.
\end{equation}
We again choose our notation such that the wavefunctions $\braket{U}{f} = f(U)$ are holomorphic in $U$ for any given vector $\ket{f}$ in the Verma module.
Note that in two dimensions the adjoint is implicitly worked into the analogue equation $\mel{u}{L_n}{f} = \mathfrak{L}_n\braket{u}{f} = \mathfrak{L}_n f(u)$ to match notation in \cite{Kraus2020}.

\section{Derivation and Solutions of Oscillator Wave Equations}
\label{sec:derivation_solution_oscillator_eqs}
Based on the oscillator representations, we now introduce the formalism we use to compute conformal blocks in both two and four dimensions. This involves \emph{oscillator wavefunctions} that can be understood as wavefunctions in the respective Bergman space, parameterised by a point $z$ in a two- or $x_\mu$ in four-dimensional Euclidean space.

\subsection{Oscillator Wave Equations in Two Dimensions}
As a next step towards the computation of conformal blocks from oscillator representations, we introduce holomorphic primary operators $\Op_{h}$ labelled by their conformal weight $h$ and defined through their action on the vacuum state $\Op_h(0)\ket{0}=\ket{h}$ at the origin\footnote{For the sake of a more transparent notation, especially when dealing with higher-point blocks, we use $h$ for the conformal weight (of an external operator) and $\h$ for the weights of the representation on the Bergman space, similar to \cite{Rosenhaus:2018zqn}.}.
Shifting the primary operator to an arbitrary point on the complex plane (more precisely, the Riemann sphere) and taking the product with a coherent state, defines the wavefunction\footnote{Note that oscillator wavefunctions in two dimensions were already derived in \cite{Kraus2020}.} 
\begin{equation}
    \psi(z;u) \equiv \psi_{h,\mathbb{h}}(z;u) = {}_{\hbb}\hspace{-.08cm}\mel{u}{\Op_{h}(z)}{0}\label{eq:SL2R_definition_psi_leg}
\end{equation}
as an element of $\BSpace$ for any given point $z$. To compute this function, we use that the commutators of the global conformal generators with primary operators can be represented as differential operators acting on the primary, i.\,e.
\begin{equation}
\label{eq:adj_gen_2d}
    [L_n,\Op_{h}(z)] = -\mathcal{L}_n\Op_{h}(z)\,, \qquad \mathcal{L}_n = -z^{n+1}\partial_z - (n+1)hz^n\,. 
\end{equation}
Note that $\mathcal{L}_n$ acts with respect to the position variable $z$ and explicitly depends on the conformal weight $h$, while $\mathfrak{L}_n$ acts with respect to the oscillator variable $u$ and depends on the Bergman weight $\h$. We sometimes write $\mathcal{L}_n^{(z)}$ and $\mathfrak{L}_n^{(u)}$ to highlight this dependency.
From equation \eqref{eq:osc_gen_2d} it is clear that these generators are adjoint to the oscillator generators, up to an extra minus sign, $\mathcal{L}_n^{(z)} = -\mathfrak{L}_{-n}^{(z)}$. Since the vacuum state satisfies $L_n\ket{0} = 0$, we obtain the linear partial differential equations
\begin{equation}
    0 = \mel{u}{\Op_{h}(z)L_n}{0} = \mel{u}{L_n\Op_{h}(z) - [L_n,\Op_{h}(z)]}{0} = (\mathfrak{L}_n + \mathcal{L}_n)\,\psi(z;u) \label{eq:SL2R_diffeq_psi_leg}
\end{equation}
for $n=0,\pm 1$. We refer to this type of equation as \emph{oscillator equation} for the oscillator wavefunction $\psi(z;u)$.  The solution to \eqref{eq:SL2R_diffeq_psi_leg} (up to a constant prefactor) is given by 
\begin{equation}
    \psi(z;u) = (1-zu)^{-2\mathbb{h}}\delta_{\mathbb{h},h}\,. \label{eq:SL2R_wave_psi_line}
\end{equation}
In similar fashion to $\psi(z;u)$, we define the anti-holomorphic wavefunction 
\begin{equation}
    \chi(z;\bar{u}) \equiv \chi_{h,\mathbb{h}}(z;\bar{u}) = \mel{0}{\Op_{h}(z)}{\bar{u}}_{\mathbb{h}} \label{eq:SL2R_definition_chi_leg}
\end{equation}
corresponding to the conjugate of a displaced primary state. Repeating the steps from above, we obtain a system of differential equations 
\begin{equation}
    0 = (-\bar{\mathfrak{L}}^{(\Bar{u})}_{-n} +\mathcal{L}^{(z)}_n)\chi(z;\bar{u}) \label{eq:SL2R_diffeq_chi_leg}
\end{equation}
that suffices to fix $\chi(z;\bar{u})$ as its solution, reading 
\begin{equation}
    \chi(z;\bar{u}) = (z-\bar{u})^{-2\mathbb{h}}\,\delta_{\mathbb{h}, h}\,. \label{eq:SL2R_wave_chi_line}
\end{equation}
Note that the wavefunction $\chi$ has the form of a conformal two-point function -- this is because its oscillator equation resembles the corresponding conformal Ward identity.

Sometimes $\psi(z;u)$ is called the in-going and $\chi(z;\Bar{u})$ the out-going wavefunction. Together, we refer to them as \emph{first-level wavefunctions} because they contain one operator insertion each. Unsurprisingly, the \emph{second-level wavefunctions}
\begin{align}
 \psi(z_1,z_2;u) \equiv \psi_{h_1,h_2,\mathbb{h}}(z_1,z_2;u) &= {}_{\mathbb{h}}\hspace{-.08cm}\mel{u}{\Op_{h_1}(z_1)\Op_{h_2}(z_2)}{0}, \label{eq:SL2R_definition_psi}\\ \chi(z_1,z_2;\bar{u}) \equiv \chi_{h_1,h_2,\mathbb{h}}(z_1,z_2;\bar{u}) &= \mel{0}{\Op_{h_1}(z_1)\Op_{h_2}(z_2)}{\bar{u}}_{\mathbb{h}}\label{eq:SL2R_definition_chi}
\end{align}
contain two primary operators inserted at the points $z_1$ and $z_2$. The additional primary compared to the first-level wavefunctions translates to an extra commutation one has to perform in the derivation of the corresponding differential equations, leading to an additional differential operator. Thus, the oscillator equations for the second-level wavefunctions read
\begin{align}
    0 &= (\mathfrak{L}_n^{(u)} + \mathcal{L}_n^{(z_1)} + \mathcal{L}_n^{(z_2)} )\,\psi(z_1,z_2;u) \label{eq:SL2R_diffeq_psi} \\
    0 &= (-\bar{\mathfrak{L}}_{-n}^{(\bar{u})} + \mathcal{L}_n^{(z_1)} + \mathcal{L}_n^{(z_2)}  )\,\chi(z_1,z_2;\bar{u}). \label{eq:SL2R_diffeq_chi}
\end{align}
Solving \eqref{eq:SL2R_diffeq_psi} and \eqref{eq:SL2R_diffeq_chi}, we find the following explicit expressions for the second-level wavefunctions
\begin{align}
    \psi(z_1,z_2;u) &= z_{12}^{\mathbb{h}-h_1-h_2}(1-z_1u)^{h_2-\mathbb{h}-h_1}(1-z_2u)^{h_1-\mathbb{h}-h_2} \label{eq:SL2R_wave_psi} \\
    \chi(z_1,z_2;\bar{u}) &= z_{12}^{\mathbb{h}-h_1-h_2}(z_1-\bar{u})^{h_2-\mathbb{h}-h_1}(z_2-\bar{u})^{h_1-\mathbb{h}-h_2} \label{eq:SL2R_wave_chi}
\end{align}
with $z_{ij} = z_i - z_j$ as usual. 

As we will see later on, these oscillator wavefunctions are the fundamental ingredients for the computation of conformal blocks in the oscillator formalism. However, to approach higher-point blocks we need one more wavefunction which is the matrix element
\begin{equation}
    \Omega(z;u_1,\bar{u}_2) \equiv \Omega_{h,\mathbb{h}_1,\mathbb{h}_2}(z;u_1,\bar{u}_2) 
    = {}_{\mathbb{h}_1} \hspace{-.1cm} \mel{u_1}{\Op_{h}(z)}{\bar{u}_2}_{\mathbb{h}_2}\,. 
\end{equation}
Once more, we can derive a set of oscillator equations  
\begin{align}
\begin{split}
    0 &= {}_{\mathbb{h}_1}\hspace{-.1cm}\mel{u_1}{L_n \Op_{h}(z) - \Op_{h}(z) L_n - [L_n,\Op_{h}(z)] }{\bar{u}_2}_{\mathbb{h}_2} \\
    &= (\mathfrak{L}_n^{(u_1)} - \bar{\mathfrak{L}}_{-n}^{(\bar{u}_2)}+\mathcal{L}_n^{(z)})\; \Omega(z;u_1,\bar{u}_2)\, \label{eq:SL2R_diffeq_omega}
    \end{split}
\end{align}
fixing $\Omega(z;u_1,\bar{u}_2)$, and obtain the result
\begin{equation}
    \Omega(z;u_1,\bar{u}_2) = (z-\bar{u}_2)^{\mathbb{h}_1-\mathbb{h}_2-h}(1-zu_1)^{\mathbb{h}_2-\mathbb{h}_1-h}(1-u_1\bar{u}_2)^{h-\mathbb{h}_1-\mathbb{h}_2}\,. \label{eq:SL2R_wave_omega}
\end{equation}

Note that the oscillator wavefunctions are not completely independent and, for example, the functional dependence of $\psi$ and $\Omega$ on their respective three arguments is the same. This observation manifests itself as the identity
\begin{equation}
    \psi_{h,\hbb_2,\hbb_1}(z,\bar{u}_2;u_1) \equiv \psi_{h,h_2,\mathbb{h}}(z,z_2;u_1)\big|_{z_2=\bar{u}_2,h_2=\hbb_2} = \Omega_{h,\mathbb{h}_1,\mathbb{h}_2}(z;u_1,\bar{u}_2) \,. \label{eq:psi_Omega_correspondence}
\end{equation}
We put this correspondence to use later on in the inductive computation of the $n$-point comb channel block.

It may be worth pointing out that one does not gain any further insight from studying higher-level oscillator equations. Starting at third-level they no longer suffice to completely fix the wavefunctions. 

For developing an intuition, it proves helpful to introduce a diagrammatic language for the wavefunctions:
\begin{gather*}
     \chi(z;\bar{u}) = 
\def\tkzscl{1}
\begin{tikzpicture}[baseline={([yshift=-1ex]current bounding box.center)},vertex/.style={anchor=base,
     circle,fill=black!25,minimum size=18pt,inner sep=2pt},scale=\tkzscl]
         \coordinate[label=left:$z$\,] (z) at (2.5,0);
         \coordinate[label=right:\,$\Bar{u}\phantom{|}$] (u) at (4,0);
         \draw[thick,dashed] (4,0) --node[above] {} (z);
         \filldraw [fill=white] (u) circle (3pt);
        \filldraw (u) circle (1pt);
        \draw (u) circle (3pt);
         \fill (z) circle (3pt);
\end{tikzpicture}
\,, \qquad \psi(z;u) = 
\def\tkzscl{1}
\begin{tikzpicture}[baseline={([yshift=-1ex]current bounding box.center)},vertex/.style={anchor=base,
     circle,fill=black!25,minimum size=18pt,inner sep=2pt},scale=\tkzscl]
         \coordinate[label=left:$\phantom{|}u$\,] (u) at (2.5,0);
         \coordinate[label=right:\,$z$] (z) at (4,0);
         \draw[thick,dashed] (4,0) --node[above] {} (u);
         \fill[white] (u) circle (3pt);
         \draw(u) circle (3pt);
         \fill (z) circle (3pt);
     \end{tikzpicture}
  \\
    \chi(z_1,z_2;\Bar{u}) = 
\def\tkzscl{0.5}
\begin{tikzpicture}[baseline={([yshift=-.5ex]current bounding box.center)},vertex/.style={anchor=base,
    circle,fill=black!25,minimum size=18pt,inner sep=2pt},scale=\tkzscl]
        \coordinate[label=left:$z_2$\,] (z_1) at (-2,2);
        \coordinate[label=left:$z_1$\,] (z_2) at (-2,-2);
        \coordinate[label=right:\,$\Bar{u}$] (u_1) at (3,0);
        \draw[thick] (z_1) -- (0,0);
        \draw[thick] (z_2) -- (0,0);
        \draw[thick, dashed] (0,0) -- node[above] {$\mathbb{h}$} (u_1);
        \fill (z_1) circle (6pt);
        \fill (z_2) circle (6pt);
        \filldraw [fill=white] (u_1) circle (6pt);
        \filldraw (u_1) circle (2pt);
        \draw (u_1) circle (6pt);
\end{tikzpicture}
\,, \qquad \psi(z_1,z_2;u) = 
\def\tkzscl{0.5}
\begin{tikzpicture}[baseline={([yshift=-.5ex]current bounding box.center)},vertex/.style={anchor=base,
    circle,fill=black!25,minimum size=18pt,inner sep=2pt},scale=\tkzscl]
        \coordinate[label=right:\,$z_1$] (z_1) at (2,2);
        \coordinate[label=right:\,$z_2$] (z_2) at (2,-2);
        \coordinate[label=left:$u$\,] (u_1) at (-3,0);
        \draw[thick] (z_1) -- (0,0);
        \draw[thick] (z_2) -- (0,0);
        \draw[thick, dashed] (0,0) -- node[above] {$\mathbb{h}$} (u_1);
        \fill (z_1) circle (6pt);
        \fill (z_2) circle (6pt);
        \fill[white] (u_1) circle (6pt);
        \draw (u_1) circle (6pt);
\end{tikzpicture}
  \\[-0.5cm]
    \Omega(z,u_1;\Bar{u}_2) = 
\def\tkzscl{0.5}
\begin{tikzpicture}[baseline={([yshift=-.5ex]current bounding box.center)},vertex/.style={anchor=base,
    circle,fill=black!25,minimum size=18pt,inner sep=2pt},scale=\tkzscl]
        \coordinate[label=left:$z$\,] (z_1) at (0,2);
        \coordinate[label=left:$u_1$\,] (u_1) at (-3,0);
        \coordinate[label=right:\,$\bar{u}_2$] (u_2) at (3,0);
        \draw[thick] (z_1) -- (0,0);
        \draw[thick, dashed] (u_2) --node[above] {$\mathbb{h}_2$} (0,0);
        \draw[thick, dashed] (0,0) -- node[above] {$\mathbb{h}_1$} (u_1);
        \fill (z_1) circle (6pt);
        \fill[white] (u_1) circle (6pt);
        \draw (u_1) circle (6pt);
        \fill[white] (u_2) circle (6pt);
        \filldraw (u_2) circle (2pt);
        \draw (u_2) circle (6pt);
        \draw[draw=none] (0,0) -- (0,-2.4);
\end{tikzpicture}
 \numberthis
\end{gather*}

\vspace{-0.9cm}
\noindent Therein, the dashed lines are labelled by the weight of the Bergman space, while solid lines correspond to the weights of the primaries. Besides the labels, these diagrams are not modified when we discuss the four-dimensional case.

\subsection{Oscillator Wave Equations in Four Dimensions}
Two-dimensional conformal field theories split into two analytically independent sectors, each respectively dependent on a complex variable $z$ and $\Bar{z}$. In four dimensions, there is no such split, so we define a CFT on four-dimensional (conformally compactified) Euclidean space with real coordinates $(x_0,x_1,x_2,x_3)$.  We map these coordinates to a complex matrix $X \in \mathbb{C}^{2\times 2}$ through the Pauli matrix basis
\begin{equation} \label{eq:X_parametr}
    X = x_0 \sigma^0 + ix_1 \sigma^1 + ix_2 \sigma^2 + i x_3 \sigma^3\,.
\end{equation}
Note the different sign convention in contrast to equation \eqref{eq:def_U_matrix}, the parametrisations of $U$ and $X$ are chosen in such a way that their contraction (modulo normalisation) gives $\delta_{\mu \nu} x^\mu u^\nu$, see e.\,g.\ \cite{Adamo:2017qyl}.

We restrict ourselves to external scalar primary operators $\mathcal{O}_\Delta(X)$ and because we also only consider the scalar contribution to the OPE, we focus on the subspace $\bigoplus_\Delta \mathcal{V}_\Delta \subset \mathcal{H}_{\text{CFT}}$. Again, operator insertions at the origin $\mathcal{O}_\Delta(0)$ create highest-weight states as $\mathcal{O}_\Delta(0) \ket{0} = \ket{\Delta}$. Conceptually, most parts of the derivations in two dimensions can then be extended to four dimensions: as in two dimensions, the holomorphic wavefunction is defined by\begin{equation}
    \psi(X;U) \equiv \psi_{\Delta,\Deltabb}(X;U) = {}_{\Deltabb}\hspace{-.08cm}\bra{U} \mathcal{O}_\Delta(X) \ket{0}\,, \label{eq:psi_4d}
\end{equation}
whereas the anti-holomorphic wavefunction is given by
\begin{equation}
    \chi(X;U^\dagger) \equiv \chi_{\Delta,\Deltabb}(X;U^\dagger) = \bra{0}\mathcal{O}_\Delta(X) \ket*{U^\dagger}_{\Deltabb}\,. \label{eq:chi_4d}
\end{equation}
Note that we have chosen our notation in a way that makes the dependence of $\chi$ on $U^\dagger$ explicit. The defining oscillator equations for \eqref{eq:psi_4d} and \eqref{eq:chi_4d} can be derived analogously to two dimensions, with the generators acting on the wavefunctions as defined in equation \eqref{eq:draw_out_generator_4d}:
\begin{align}
\label{eq:osc_eq_psi_4d}
    \left(-\mathcal{L}^{(x)}+{\mathfrak{L}^{(w)\dagger}}\right)\psi(x;w) &= 0\\
    \left(\mathcal{L}^{(x)} + \mathfrak{L}^{(\bar{w})}\right)  \chi(x;\bar{w})& = 0\,.
\end{align}
Here we are using the coordinate notation introduced in section \ref{ssec:4d_Oscillator}, as the differential op\-er\-ators are given in terms of the coordinates $w_\mu$ and $x_\mu$ instead of the corresponding matrices $U$ and $X$. In the adjoint representation, the generators $\mathcal{L}^{(x)}$ acting with respect to the coordinates $x_\mu$ are given by 
\begin{align}
\begin{split}
    \mathcal{P}_\mu &= \partial_\mu\,,\quad \mathcal{K}_\mu = x^2\partial_\mu-2x_\mu x^\rho\partial_\rho-2x_\mu\Delta\,,\\
    \mathcal{D} &= x^\mu \partial_\mu + \Delta\,,
    \quad
    \mathcal{M}_{\mu\nu} = x_\mu\p_\nu-x_\nu\p_\mu\,.
\end{split}
\end{align}
Now one finds that for the wavefunction $\chi(x;\bar{w})$ the oscillator equations read 
\begin{subequations}
\label{eq:Ward}
\begin{align}
    \left( \mathcal{D}^{(x)} + \mathfrak{D}^{(\bar{w})} \right) \chi(x;\bar{w}) &= 0 \, , \\ 
    \left( \mathcal{P}_\mu^{(x)} + \mathfrak{P}_\mu^{(\bar{w})} \right) \chi(x;\bar{w}) &= 0 \, , \\ 
    \left( \mathcal{K}_\mu^{(x)} + \mathfrak{K}_{\mu}^{(\bar{w})} \right) \chi(x;\bar{w}) &= 0 \, , \\ 
    \left( \mathcal{M}_{\mu\nu}^{(x)} + \mathfrak{M}_{\mu\nu}^{(\bar{w})} \right) \chi(x;\bar{w}) &= 0 \, .
\end{align}
\end{subequations}
Similar to two dimensions, they resemble the conformal Ward identities, such that the solution can be directly found as
\begin{align}
    \chi(x;\bar{w}) &=  \left((x-\bar{w})^2\right)^{-\Deltabb}\delta_{\Delta,\Deltabb}\,.
\end{align}
This expression can be translated back to the matrix notation by a straightforward calculation, yielding
\begin{align}  \chi(X;U^\dagger) = \mathrm{det}^{-\Deltabb} (X-U^\dagger)\,\delta_{\Delta,\Deltabb}\,. \label{eq:chi_4d_matrix}
\end{align}
Accordingly, for the holomorphic wavefunction $\psi(x;w)$ one gets the following set of equations:
\begin{subequations}
\label{eq:deq_psi}
\begin{align}
    \left(-\mathcal{D}^{(x)} + \mathfrak{D}^{(w)} \right) \psi(x;w) &= 0 \, , \\ 
    \left( \mathcal{P}_\mu^{(x)} + \mathfrak{K}_\mu^{(w)} \right) \psi(x;w) &= 0 \, , \\ 
     \left( \mathcal{K}_\mu^{(x)} + \mathfrak{P}_\mu^{(w)} \right) \psi(x;w) &= 0 \, , \\
    \left( \mathcal{M}_{\mu\nu}^{(x)} + \mathfrak{M}_{\mu\nu}^{(w)} \right)\psi(x;w) &= 0 \, ,
\end{align}
\end{subequations}
solved by 
\begin{align}
    \psi(x;w) &= \left(1 - 2\,w \cdot x +w^2 x^2\right)^{-\Deltabb }\delta_{\Delta,\Deltabb} = \mathrm{det}^{-\Deltabb}(1-UX)\,\delta_{\Delta,\Deltabb}\,, \label{eq:4d_psi_solution}
\end{align}
with $w\cdot x=w_\mu x^\mu$. We evidently find that the first-level wavefunctions in \eqref{eq:chi_4d_matrix} and \eqref{eq:4d_psi_solution} have the same structure as the Bergman kernel in \eqref{eq:expansion_kernel}, as was the case in two dimensions.

We continue by defining second-level wavefunctions featuring two primary operator insertions
\begin{align}     
    \chi(X_1,X_2;U^\dagger)&\equiv\chi_{\Delta_1,\Delta_2,\Deltabb}(X_1,X_2;U^\dagger) = \mel*{0}{\mathcal{O}_{\Delta_1}(X_1)\mathcal{O}_{\Delta_2}(X_2)}{U^\dagger}_{\Deltabb}\,,\label{eq:2ndlevel_chi_osc_eq}\\
    \psi\left(X_1,X_2;U\right)&\equiv\psi_{\Delta_1,\Delta_2,\Deltabb}\left(X_1,X_2;U\right) = {}_\Deltabb\hspace{-.08cm}\mel{U}{\mathcal{O}_{\Delta_1}(X_1)\mathcal{O}_{\Delta_2}(X_2)}{0}\,,
\end{align}
as well as the matrix element $\Omega(X;U_1,U_2^{\dagger})$ depending on two oscillator variables
\begin{equation}
\Omega(X;U_1,U_2^{\dagger}) \equiv \Omega_{\Delta,\mathbb{\Delta}_1,\mathbb{\Delta}_2}(X;U_1,U_2^{\dagger}) = {}_{\Deltabb_1}\hspace{-.08cm}{\mel*{U_1}{\Op_{\Delta}(X)}{U_2^{\dagger}}_{\Deltabb_2}}\,. 
\end{equation}
These obey oscillator equations with three differential generators, namely:
\begin{align}
    0 &= \left(\mathfrak{L}^{(\bar{w})\phantom{\dagger}}\!\! + \mathcal{L}^{(x_1)} + \mathcal{L}^{(x_2)}\right)\chi(x_1,x_2;\bar{w})\,, \label{eq:4d_diffeq_chi_2nd}\\
    0 &= \left(\mathfrak{L}^{(w)\dagger} - \mathcal{L}^{(x_1)} - \mathcal{L}^{(x_2)} \right)\psi(x_1,x_2;w)\,, \label{eq:4d_diffeq_psi_2nd} \\
    0 &= \left(\mathfrak{L}^{(w_1)\dagger} - \mathfrak{L}^{(\bar{w}_2)}-\mathcal{L}^{(x)}\right) \Omega(x;w_1,\bar{w}_2)\,. \label{eq:4d_diffeq_omega_2nd}
\end{align}
The results of the corresponding oscillator equations are given by
\begin{align}
    \chi(X_1,X_2;U^\dagger)&=  \mathrm{det}^{-\alpha}(X_1-U^\dagger)\, \mathrm{det}^{-\beta}(X_2-U^\dagger)\,\mathrm{det}^{-\gamma}(X_1-X_2)\,,\label{eq:sol_chi_2ndLvl_V1}\\
    \psi(X_1,X_2;U) &= \mathrm{det}^{-\alpha}\left(\mathbb{1} -U {X_1} \right) \mathrm{det}^{-\beta}\left(\mathbb{1} -U {X_2}\right)\mathrm{det}^{-\gamma}(X_1-X_2)\,, \label{eq:sol_psi_2ndLvl_V1}
\end{align}
with exponents  
\begin{align}
    \alpha &= \frac{1}{2}(\Delta_1+\Deltabb-\Delta_2)\,, & \beta &= \frac{1}{2}(\Delta_2+\Deltabb-\Delta_1)\,, & \gamma &= \frac{1}{2}(\Delta_1+\Delta_2-\Deltabb)\,. 
\end{align}
Much like in two dimensions, the function $\Omega$ can be retrieved by $\Omega(X;U_1,U^\dagger_2) = \psi(X,U^\dagger_2;U_1)$, yielding
\begin{align}
    \Omega(X;U_1,U_2^{\dagger}) &=  \mathrm{det}^{-\Tilde{\alpha}}(\mathbb{1} -U_1 X) \text{det}^{-\Tilde{\beta}}(\mathbb{1} -U_1 {U_2^{\dagger}})\, \text{det}^{-\Tilde{\gamma}}(X-U_2^{\dagger})\,, \label{eq:sol_Omega_2ndLvl_V1}
\end{align}
where the modified exponents read:
\begin{align}
     \Tilde{\alpha} &= \frac{1}{2}(\Delta+\Deltabb_1-\Deltabb_2)\,,  & \Tilde{\beta} &= \frac{1}{2}(\Deltabb_2+\Deltabb_1-\Delta)\,,& \Tilde{\gamma} &= \frac{1}{2}(\Delta+\Deltabb_2-\Deltabb_1)\,. 
\end{align}
We have now gathered all the necessary ingredients for the computation of conformal blocks and start with lower-point correlation functions.

\section{Two- and Three-Point Correlation Functions}
\label{sec:two_three_point_correlation_fct}
To demonstrate the formalism, we give the step-by-step calculation of the easiest non-trivial cases: the two- and three-point correlation functions. We compute them by inserting a complete set of states of the CFT Hilbert space $\mathcal{H}_{\text{CFT}}$ into the correlation functions to obtain them as integrals of the previously discussed wavefunctions.

\subsection{Two- and Three-Point Functions in Two Dimensions}
By definition, $\mathbb{P}_\hbb$ projects from the holomorphic sector of the CFT Hilbert space \mbox{$\mathcal{H}_{\text{CFT}}\! =\! \bigoplus_\hbb\!\! \mathcal{V}_\hbb$} onto one specific Verma module $\mathcal{V}_\hbb$, and thus $\mathbb{1}=\sum_\hbb\mathbb{P}_\hbb$.
The insertion of a projector $\mathbb{P}_\hbb$ on $\mathcal{H}L^2_\hbb(\mathbb{D})$ into the two-point function can be thought of as placing $\mathbb{P}_\hbb$ as a surface operator on the circle that separates the two points $z_1$ and $z_2$ \cite{Simmons-Duffin:2012juh, Simmons-Duffin:2016gjk}. The integral representation of the projector on the Bergman space gives us an expression involving the aforementioned first-level wavefunctions:
\begin{align}
\begin{split}
    \expval{\mathcal{O}_{h_1}(z_1) \mathcal{O}_{h_2}(z_2)} &=\sum\limits_\hbb\expval{\mathcal{O}_{h_1}(z_1) \mathbb{P}_\hbb\, \mathcal{O}_{h_2}(z_2)}\\
    &= \sum\limits_\hbb\int_\mathbb{D} [\d u]_\hbb \mel{0}{\mathcal{O}_{h_1}(z_1)}{\Bar{u}}\!\! \mel{u}{\mathcal{O}_{h_2}(z_2)}{0} \\ 
    &= \sum\limits_\hbb\int_\mathbb{D} [\d u]_\hbb\, \chi(z_1;\Bar{u}) \psi(z_2;u) \\
    &= \sum\limits_\hbb\int_\mathbb{D} [\d u]_\hbb\, (z_1-\Bar{u})^{-2\hbb} (1-z_2u)^{-2\hbb} \delta_{h_1,\hbb}\, \delta_{\hbb,h_2}\,.
\end{split}
\end{align}
As expected, we only obtain a non-vanishing two-point function for $h_1=h_2$. By expanding each wavefunction into monomials through the binomial theorem and using the inner product \eqref{eq:2d_orthogonality_relation}, we find 
\begin{align}
    \expval{\mathcal{O}_{h_1}(z_1) \mathcal{O}_{h_1}(z_2)}  &= z_1^{-2h_1} \sum_{m,n=0}^\infty (-1)^{m+n}\binom{-2h_1}{m} \binom{-2h_1}{n} z_1^{-m} z_2^n \int_\mathbb{D} [\d u]_{h_1} \Bar{u}^m u^n\,\delta_{h_1,h_2}\notag\\
    &= \sum_{m=0}^\infty \binom{-2h_1}{m} \binom{-2h_1}{m} z_1^{-2h_1-m} z_2^m \frac{m!}{(2h_1)_m}\delta_{h_1,h_2} \\
    &= (z_1-z_2)^{-2h_1}\delta_{h_1,h_2}\,,\notag
\end{align}
where in the last step we rewrote the binomial coefficients in terms of Pochhammer symbols and resummed using the binomial theorem. Note that diagrammatically we have just computed 
\begin{equation}
    \sum\limits_\hbb\int_{\mathbb{D}} [\d u]_\hbb \def\tkzscl{1}
\begin{tikzpicture}[baseline={([yshift=-1ex]current bounding box.center)},vertex/.style={anchor=base,
     circle,fill=black!25,minimum size=18pt,inner sep=2pt},scale=\tkzscl]
         \coordinate[label=left:$\phantom{|}z_1$\,] (z) at (2.5,0);
         \coordinate[label=right:\,$\Bar{u}\phantom{|}$] (u) at (4,0);
         \draw[thick,dashed] (4,0) --node[above] {} (z);
         \filldraw [fill=white] (u) circle (3pt);
        \filldraw (u) circle (1pt);
        \draw (u) circle (3pt);
         \fill (z) circle (3pt);
\end{tikzpicture}  \def\tkzscl{1}
\begin{tikzpicture}[baseline={([yshift=-1ex]current bounding box.center)},vertex/.style={anchor=base,
     circle,fill=black!25,minimum size=18pt,inner sep=2pt},scale=\tkzscl]
         \coordinate[label=left:$\phantom{|}u$\,] (u) at (2.5,0);
         \coordinate[label=right:\,$z_2\phantom{|}$] (z) at (4,0);
         \draw[thick,dashed] (4,0) --node[above] {} (u);
         \fill[white] (u) circle (3pt);
         \draw(u) circle (3pt);
         \fill (z) circle (3pt);
     \end{tikzpicture}
    = 
\def\tkzscl{1}
\begin{tikzpicture}[baseline={([yshift=-1ex]current bounding box.center)},vertex/.style={anchor=base,
     circle,fill=black!25,minimum size=18pt,inner sep=2pt},scale=\tkzscl]
         \coordinate[label=left:$\phantom{|}z_1$\,] (u) at (2.5,0);
         \coordinate[label=right:\,$z_2\phantom{|}$] (z) at (4,0);
         \draw[thick] (4,0) --node[above] {} (u);
         \fill (u) circle (3pt);
         \fill (z) circle (3pt);
     \end{tikzpicture}
\end{equation}

In the case of the three-point function, we have some freedom in where to insert the projector. The following choice gives
\begin{align}
 \expval{\mathcal{O}_{h_1}\!(z_1)\mathcal{O}_{h_2}\!(z_2) \mathcal{O}_{h_3}\!(z_3)}\!&=\sum\limits_\hbb\expval{\mathcal{O}_{h_1}(z_1)\mathcal{O}_{h_2}(z_2) \mathbb{P}_\hbb\mathcal{O}_{h_3}(z_3)}  \notag\\ 
 \begin{split}
 &= \sum\limits_\hbb\int_{\mathbb{D}} [\d u]_\hbb \mel{0}{\mathcal{O}_{h_1}(z_1)\mathcal{O}_{h_2}(z_2)}{\bar{u}}\!\!\mel{u}{\mathcal{O}_{h_3}(z_3)}{0}\\
&= \sum\limits_\hbb\int_{\mathbb{D}} [\d u]_\hbb\, \chi(z_1,z_2;\bar{u}) \psi(z_3; u)
\end{split}\\
&= z_{12}^{h_3 - h_2 - h_1} \!\!\int_{\mathbb{D}} [\d u]_{h_3} (z_1-\bar{u})^{h_2 - h_3 - h_1} (z_2-\bar{u})^{h_1 - h_3 - h_2} (1-z_3 u)^{-2h_3}\,. \notag   
\end{align}
In the last step, the sum has been evaluated as the first level wavefunction comes with $\delta_{h_3,\hbb}$. Expanding this again in monomials and using once more the orthogonality relation \eqref{eq:2d_orthogonality_relation}, we find the expected result
\begin{equation}
\label{eq:2d_3ptfct}
    \expval{\mathcal{O}_{h_1}(z_1)\mathcal{O}_{h_2}(z_2) \mathcal{O}_{h_3}(z_3)} =  z_{12}^{-h_1 - h_2 +h_3} z_{13}^{-h_1-h_3 +h_2} z_{23}^{-h_2 - h_3 +h_1}\,.   
\end{equation}
One can easily check that all other possible projector insertions give the same result.

\subsection{Two- and Three-Point Functions in Four Dimensions}
The two-point correlation function of scalar operators is computed conceptually in the same way as in two dimensions, but now we insert a projector $\mathbb{P}_{\Deltabb}$ on the sphere that separates $X_1$ from $X_2$, which leads to 
\begin{align}
\begin{split}
    \expval{\mathcal{O}_{\Delta_1}(X_1) \mathcal{O}_{\Delta_2}(X_2)}&=\sum\limits_\Deltabb\expval{\mathcal{O}_{\Delta_1}(X_1) \mathbb{P}_\Deltabb \mathcal{O}_{\Delta_2}(X_2)}\\
    &=\sum\limits_\Deltabb\int_{\Disc_4} [\dd U]_\Deltabb\, \mel*{0}{\mathcal{O}_{\Delta_1}(X_1)}{U^\dagger}\mel{U}{\mathcal{O}_{\Delta_2}(X_2)}{0}\\
    &=\sum\limits_\Deltabb\int_{\Disc_4} [\dd U]_\Deltabb\, \mathrm{det}^{-\Delta_1}(X_1-U^\dagger)\,\mathrm{det}^{-\Delta_2}(\mathbb{1}-X_2U)\, \delta_{\Delta_1,\Deltabb} \delta_{\Delta_2,\Deltabb} \,,
    \end{split}
    \end{align}
where we used the matrix notation for operator insertions as in equation \eqref{eq:X_parametr}. While in two dimensions the next step was an expansion in monomials, here, we expand the determinants analogously to the kernel in \eqref{eq:expansion_kernel}. We explain at the end of section \ref{sec:conf_blocks_comb_channel} why this is applicable. 
After expanding, we evaluate the sum over $\Deltabb$ such that both Kronecker deltas reduce to $\delta_{\Delta_1,\Delta_2}$, yielding 
    \begin{align}
    &\expval{\mathcal{O}_{\Delta_1}(X_1) \mathcal{O}_{\Delta_2}(X_2)}\\
    &= x_1^{-2\Delta_1}\sum\limits_{q_a,q_b}^{j,m} \sum\limits_{q'_a,q'_b}^{j',m'} \left(\mathcal{N}^{j,m}_{\Delta_1} \mathcal{N}^{j',m'}_{\Delta_2}\right)^2 \phi_{q_a,q_b}^{j,m}(X_1^{-1}) \phi_{q'_a,q'_b}^{j',m'}(X_2^{T})
     \int_{\Disc_4}[\dd U]_{\Delta_1} \overline{\phi_{q_a,q_b}^{j,m}(U)} \phi_{q'_a,q'_b}^{j',m'}(U)\delta_{\Delta_1,\Delta_2}\,.\notag
     \end{align}
Next, we use the inner product defined in equation \eqref{eq:4d_orthonormality_property} to perform the integral
\begin{align}
\begin{split}
    &\expval{\mathcal{O}_{\Delta_1}(X_1) \mathcal{O}_{\Delta_2}(X_2)}
    \\
    &= x_1^{-2\Delta_1}\sum\limits_{q_a,q_b}^{j,m}  \sum\limits_{q'_a,q'_b}^{j',m'} \frac{\left(\mathcal{N}^{j,m}_{\Delta_1}\right)^2 \left(\mathcal{N}^{j',m'}_{\Delta_1}\right)^2}{\left(\mathcal{N}^{j,m}_{\Delta_1}\right)^2} \phi_{q_a,q_b}^{j,m}(X_1^{-1})\phi_{q'_a,q'_b}^{j',m'}(X_2^{T}) \delta^{j',m',q_a',q_b'}_{j,m,q_a,q_b} \delta_{\Delta_1,\Delta_2}\\
    &= x_1^{-2\Delta_1}\sum\limits_{q_a,q_b}^{j,m} \left(\mathcal{N}^{j,m}_{\Delta_1}\right)^2 \phi_{q_a,q_b}^{j,m}(X_1^{-1})\phi_{q_a,q_b}^{j,m}(X_2^{T})\delta_{\Delta_1,\Delta_2}\\
    &=\mathrm{det}^{-\Delta_1}(X_1-X_2)\,\delta_{\Delta_1,\Delta_2}=\frac{1}{(x_1-x_2)^{2\Delta_1}}\delta_{\Delta_1,\Delta_2}\,.
    \end{split}
\end{align}
In the second-to-last step we again used equation \eqref{eq:expansion_kernel} as well as the identity $\mathcal{D}^j_{q_a,q_b}\big(X^T\big) = \mathcal{D}^j_{q_b,q_a}(X)$ for the Wigner $\mathcal{D}$-matrices. 

Following the recipe from two dimensions for the computation of the three-point function leads to a somewhat involved (but possible) calculation, which can be facilitated using translational invariance:
\begin{align}
        \expval{\mathcal{O}_{\Delta_1}(X_1)\mathcal{O}_{\Delta_2}(X_2) \mathcal{O}_{\Delta_3}(X_3) } &= \expval{\mathcal{O}_{\Delta_1}(X_1-X_2)\mathcal{O}_{\Delta_2}(0) \mathcal{O}_{\Delta_3}(X_3-X_2) }\nonumber \\
    &\equiv \expval{\mathcal{O}_{\Delta_1}(\widetilde{X}_1)\mathcal{O}_{\Delta_2}(0) \mathcal{O}_{\Delta_3}(\widetilde{X}_3) }\,. \\
    \intertext{Inserting a projector into this expression gives rise to an integral over wavefunctions}
    \expval{\mathcal{O}_{\Delta_1}(\widetilde{X}_1) \mathcal{O}_{\Delta_2}(0) \mathcal{O}_{\Delta_3}(\widetilde{X}_3) }&=\sum\limits_\Deltabb\expval{\mathcal{O}_{\Delta_1}(\widetilde{X}_1) \mathbb{P}_{\mathbb{\Delta}} \mathcal{O}_{\Delta_2}(0) \mathcal{O}_{\Delta_3}(\widetilde{X}_3) }\nonumber\\
    &= \sum\limits_\Deltabb\int_{\mathbb{D}_4} [\d U]_\Deltabb\, \chi(\widetilde{X}_1;U^\dagger)\, \psi(0,\widetilde{X}_3;U)\,,
\end{align}
where these wavefunctions evaluated at the suggested points are given by
\begin{align}
    \psi(0,\widetilde{X}_3;U) &= \text{det}^{-\beta}(\mathbb{1}-\widetilde{X}_3 U)\, \text{det}^{-\gamma}(-\widetilde{X}_3)\,,  \\
    \chi(\widetilde{X}_1;U^\dagger) &= \Tilde{x}_1^{-2\Delta_1}\,\text{det}^{-\Delta_1} (\mathbb{1}-U^\dagger \widetilde{X}_1^{-1})\, \delta_{\Delta_1,\mathbb{\Delta}}\,.
\end{align}
By exploiting translational invariance, we averted the appearance of two determinants depending on $U$ in the wavefunction $\psi$.

We calculate the three-point function by expanding the wavefunctions in the same way as before and performing the sum over $\Deltabb$:
\begin{align}
    &\sum\limits_\Deltabb\int_{\mathbb{D}_4} [\d U]_\Deltabb\,\chi(\widetilde{X}_1;U^\dagger)\, \psi(0,\widetilde{X}_3;U) \nonumber\\ 
    &= \sum\limits_\Deltabb\Tilde{x}_1^{-2\Delta_1} (-\widetilde{x}_3)^{-2\gamma}  \int_{\mathbb{D}_4} [\d U]_{\mathbb{\Delta}}\, \text{det}^{-\beta}(\mathbb{1}-\widetilde{X}_3 U)\, \text{det}^{-\Delta_1}(\mathbb{1}-U^\dagger \widetilde{X}_1^{-1})\, \delta_{\Delta_1,\mathbb{\Delta}} \\
    &= \Tilde{x}_1^{-2\Delta_1} (-\Tilde{x}_3)^{-2 \gamma} \!\sum\limits_{q_a,q_b}^{j,m} \sum\limits_{q'_a,q'_b}^{j',m'}\! \left(\mathcal{N}^{j,m}_{\beta} \mathcal{N}^{j',m'}_{\Delta_1}\right)^2 \!\!\phi_{q_a,q_b}^{j,m}(\widetilde{X}_3^{T}) \phi_{q'_a,q'_b}^{j',m'} (\widetilde{X}_1^{-1})\!\int_{\mathbb{D}_4}\!\! [\d U]_{\Delta_1} \phi_{q_a,q_b}^{j,m} (U)\,\overline{  \phi_{q'_a,q'_b}^{j',m'}(U)} \,.   \nonumber
\end{align}
Now by evaluating the integral, we can use the orthogonality relation \eqref{eq:4d_orthonormality_property} to perform the sum over the primed quantities and, with $\Delta_1 = \alpha + \beta$, arrive at
\begin{align}
    \expval{\mathcal{O}_{\Delta_1}(X_1)\mathcal{O}_{\Delta_2}(X_2) \mathcal{O}_{\Delta_3}(X_3) }
    &= \Tilde{x}_1^{-2\alpha}  (-\Tilde{x}_3)^{-2 \gamma} \Tilde{x}_1^{-2 \beta} \sum\limits_{q_a,q_b}^{j,m} \left(\mathcal{N}^{j,m}_{\beta}\right)^2\! \phi_{q_a,q_b}^{j,m} (\widetilde{X}_3^{T}) \phi_{q_a,q_b}^{j,m} (\widetilde{X}_1^{-1})\, \nonumber\\
    &= \text{det}^{-\alpha}(X_1 - X_2)\, \text{det}^{-\beta}(X_1 - X_3)\, \text{det}^{-\gamma}(X_2 - X_3) 
\end{align}
where we apply equation \eqref{eq:expansion_kernel} for the resummation. By rewriting the determinants in coordinate notation, one arrives at the well-known result.

\section{Conformal Blocks in the Comb Channel}
\label{sec:conf_blocks_comb_channel}

So far we have computed the lower-point correlation functions to familiarise ourselves with the formalism. We finally turn to the actual objects of interest: the conformal blocks. In two dimensions, we start with demonstrating how to compute the easiest case, the four-point block, and then derive the general $n$-point block in the comb channel up to technical details, which are laid out in appendix \ref{app:comb_npt}. We especially highlight the modular nature of the construction. 
In four dimensions, we calculate the scalar four-point block with scalar exchange as a proof of concept.

\subsection{Conformal Blocks in Two Dimensions}
In the case of the four-point block, we choose our cross-ratio as
\begin{equation}
    \xi = \frac{z_{12} z_{34}}{z_{13} z_{24}}\,,
\end{equation}
such that for the usual choice of coordinates $z_1 \to \infty$, $z_2 = 1$ and $z_4 = 0$, we find $\xi = z_3$. Inserting a specific projector $\mathbb{P}_\hbb$ allows us to write
\begin{equation}
    \label{eq:4pt_proj}
    \expval{\mathcal{O}_{h_1}(\infty)\mathcal{O}_{h_2} (1) \mathbb{P}_\hbb \mathcal{O}_{h_3}(\xi) \mathcal{O}_{h_4}(0)}  = \int_{\mathbb{D}} [\d u]_\hbb\, \chi(\infty,1;\Bar{u}) \psi(\xi,0;u)\,.
\end{equation}
Note that taking the limit $z_1 \to \infty$ for $\chi(z_1,z_2;\Bar{u})$ should be understood as only considering the coefficient of the leading divergent $z_1^{-2h_1}$ factor\footnote{This is exactly the factor that one gets by writing $\chi$ in terms of $\psi$, i.\,e.\ for the first-level wavefunctions $\chi(z;\Bar{u}) = z^{-2h} \psi(z^{-1},\Bar{u})$.}, such that one gets, up to constant prefactor,
\begin{align}
    \chi(\infty,1;\Bar{u}) &= (1-\Bar{u})^{h_1 - h_2 - \hbb}\,, \\
    \psi(\xi,0;u) &= \xi^{\hbb - h_4 - h_3} (1-\xi u)^{h_4 - h_3 - \hbb}\,.
\end{align}
Evaluating equation \eqref{eq:4pt_proj} in the same way as the three-point function gives the expected result of 
\begin{equation}
     \expval{\mathcal{O}_{h_1}(\infty)\mathcal{O}_{h_2} (1) \mathbb{P}_\hbb \mathcal{O}_{h_3}(\xi) \mathcal{O}_{h_4}(0)} = \xi^{\hbb-h_4-h_3} {}_2F_1\left[\begin{matrix} \hbb + h_2 - h_1, \hbb + h_3 - h_4\\ 2\hbb 
     \end{matrix}; \xi\right]\,.
\end{equation}

The four-point block has already been computed in \cite{Kraus2020} using this method; 
next we compute the higher-point blocks in the same straightforward way. For five points in two dimensions, there exist two independent cross-ratios. We choose to work with 
\begin{equation}
    \xi_1= \frac{z_{12}z_{35}}{z_{13}z_{25}}\quad \text{and} \quad \xi_2= \frac{z_{12}z_{45}}{z_{14}z_{25}}\,,
\end{equation}
as well as the point configuration $z_1 \to \infty$, $z_2 = 1$, $z_5 = 0$, in which the cross-ratios simplify to $\xi_1=z_3$ and $\xi_2 = z_4$. 
To obtain the five-point block, we need to insert projectors between the second and third as well as the third and fourth primary, i.\,e. 
\begin{align}
\begin{split}
    &\expval{\mathcal{O}_{h_1}(\infty)\mathcal{O}_{h_2} (1) \mathbb{P}_{\hbb_1} \mathcal{O}_{h_3}(\xi_1) \mathbb{P}_{\hbb_2}\mathcal{O}_{h_4}(\xi_2)\mathcal{O}_{h_5}(0)} \\
    &~~= \int_\mathbb{D}[\d u_1]_{\mathbb{h}_1}\int_\mathbb{D}[\d u_2]_{\mathbb{h}_2}\, \chi(\infty,1;\bar{u}_1)\, \Omega(\xi_1;u_1,\bar{u}_2)\, \psi(\xi_2,0;u_2)\,. \label{eq:2d_5pt_block_WFlevel}
\end{split}
\end{align}
Note how, compared to the four-point block, the matrix element $\Omega$ introduced previously comes into play allowing us to write the five-point block diagrammatically as 
\begin{equation}
    \int_{\mathbb{D}}[\d u_1]_{\hbb_1} 
\int_{\mathbb{D}}[\d u_2]_{\hbb_2}\mkern-18mu \! \!  
\def\tkzscl{0.5}
\begin{tikzpicture}[baseline={([yshift=-.5ex]current bounding box.center)},vertex/.style={anchor=base,
    circle,fill=black!25,minimum size=18pt,inner sep=2pt},scale=\tkzscl]
        \coordinate[label=left:$1$\,] (z_1) at (-2,2);
        \coordinate[label=left:$\infty$\,] (z_2) at (-2,-2);
        \coordinate[label=above:\,$\Bar{u}_1\phantom{|}$] (u_1) at (3,0);
        \draw[thick] (z_1) -- (0,0);
        \draw[thick] (z_2) -- (0,0);
        \draw[thick, dashed] (0,0) -- node[above] {$\mathbb{h}_1$} (u_1);
        \fill (z_1) circle (6pt);
        \fill (z_2) circle (6pt);
        \filldraw [fill=white] (u_1) circle (6pt);
        \filldraw (u_1) circle (2pt);
        \draw (u_1) circle (6pt);
\end{tikzpicture}
\def\tkzscl{0.5}
\begin{tikzpicture}[baseline={([yshift=-.5ex]current bounding box.center)},vertex/.style={anchor=base,
    circle,fill=black!25,minimum size=18pt,inner sep=2pt},scale=\tkzscl]
        \coordinate[label=left:$\xi_1$\,] (z_1) at (0,2.2);
        \coordinate[label=below:$u_1\phantom{|}$\,] (u_1) at (-3,0);
        \coordinate[label=below:$\phantom{|}\bar{u}_2$] (u_2) at (3,0);
        \draw[thick] (z_1) -- (0,0);
        \draw[thick, dashed] (u_2) --node[above] {$\mathbb{h}_2$} (0,0);
        \draw[thick, dashed] (0,0) -- node[above] {$\mathbb{h}_1$} (u_1);
        \fill (z_1) circle (6pt);
        \fill[white] (u_1) circle (6pt);
        \draw (u_1) circle (6pt);
        \fill[white] (u_2) circle (6pt);
        \filldraw (u_2) circle (2pt);
        \draw (u_2) circle (6pt);
        \draw[draw=none] (0,0) -- (0,-2.6);
\end{tikzpicture}
\def\tkzscl{0.5}
\begin{tikzpicture}[baseline={([yshift=-.5ex]current bounding box.center)},vertex/.style={anchor=base,
    circle,fill=black!25,minimum size=18pt,inner sep=2pt},scale=\tkzscl]
        \coordinate[label=right:\,$\xi_2$] (z_1) at (2,2);
        \coordinate[label=right:\,$0$] (z_2) at (2,-2);
        \coordinate[label=above:$u_2\phantom{|}$\,] (u_1) at (-3,0);
        \draw[thick] (z_1) -- (0,0);
        \draw[thick] (z_2) -- (0,0);
        \draw[thick, dashed] (0,0) -- node[above] {$\mathbb{h}_2$} (u_1);
        \fill (z_1) circle (6pt);
        \fill (z_2) circle (6pt);
        \fill[white] (u_1) circle (6pt);
        \draw (u_1) circle (6pt);
\end{tikzpicture}
\end{equation}

Conceptually, the computation of \eqref{eq:2d_5pt_block_WFlevel} runs exactly as was presented in detail for the lower-point correlators: We expand the wavefunctions \eqref{eq:SL2R_wave_psi}, \eqref{eq:SL2R_wave_chi} and \eqref{eq:SL2R_wave_omega} in monomials of the respective oscillator variables and use the orthogonality relation \eqref{eq:2d_orthogonality_relation} to perform the integrals. After some elementary manipulations of the remaining sums and stripping away the residue of the leg factor\footnote{Since we are mainly interested in the bare block, we divide by the leg factor that does not depend on the inner weights and, thus, is common to all five-point blocks. For the sake of comparison we explicitly state that the leg factor used in this calculation reads $z_{12}^{-h_1-h_2+h_3+h_4+h_5}z_{13}^{-2h_3}z_{14}^{-2h_4}z_{15}^{-h_1+h_2+h_3+h_4-h_5}z_{25}^{h_1-h_2-h_3-h_4-h_5}$.}, we arrive at 
\begin{align}
\begin{split}
   &\expval{\mathcal{O}_{h_1}(\infty)\mathcal{O}_{h_2} (1) \mathbb{P}_{\hbb_1} \mathcal{O}_{h_3}(\xi_1) \mathbb{P}_{\hbb_2}\mathcal{O}_{h_4}(\xi_2)\mathcal{O}_{h_5}(0)} \\
    &\quad= \xi_1^{\mathbb{h}_1-\mathbb{h}_2-h_3}\xi_2^{\mathbb{h}_2-h_4-h_5} \sum\limits_{k,l=0}^\infty \tau_{kl}\, \xi_1^{k} \left(\frac{\xi_2}{\xi_1}\right)^{l} 
    \frac{(\mathbb{h}_1-h_1+h_2)_{k}(\mathbb{h}_2+h_4-h_5)_{l}}{(2\mathbb{h}_1)_{k}(2\mathbb{h}_2)_{l}}
\end{split}
\end{align}
with coefficients 
\begin{equation}
    \tau_{kl} =  \sum\limits_{m=0}^{\min(k,l)} \frac{(\mathbb{h}_1+\mathbb{h}_2-h_3)_{m} (\mathbb{h}_1-\mathbb{h}_2+h_3)_{k-m}(\mathbb{h}_2-\mathbb{h}_1+h_3)_{l-m} }{m!(k-m)!(l-m)!}\,. \label{eq:tau_coeffs}
\end{equation}
This result agrees with the five-point block presented in \cite{Alkalaev_2016}.\footnotemark

Let us now approach the general $n$-point comb channel block, first computed in \cite{Rosenhaus:2018zqn}. For any number $n$ of external primaries, the comb block can be obtained by inserting $(n-3)$ projectors sequentially between intermediate pairs of operators, excluding the first and last pairs. This leads to the following integral expression
\begin{align}
    & G^{h_1,\dots,h_n}_{\mathbb{h}_1,\dots,\mathbb{h}_{n-3}}(z_1,\dots,z_n)  \label{eq:2d_npt_block_WFlevel} \\
    &= \expval{\Op_{h_1}(z_1) \Op_{h_2}(z_2) \mathbb{P}_{\h_1} \Op_{h_3}(z_3) \mathbb{P}_{\h_2} \dots \mathbb{P}_{\h_{n-4}} \Op_{h_{n-2}}(z_{n-2}) \mathbb{P}_{\h_{n-3}} \Op_{h_{n-1}}(z_{n-1}) \Op_{h_{n}}(z_n) }  \nonumber\\
    &=\!\int_\mathbb{D}\![\d u_1]_{\hbb_1}\!\dots\! \int_\mathbb{D}\![\d u_{n-3}]_{\hbb_{n-3}} \chi(z_1,z_2;\bar{u}_1)\Omega(z_3;u_1,\bar{u}_2)\dots \Omega(z_{n-2};u_{n-4},\bar{u}_{n-3}) \psi(z_{n-1},z_n;u_{n-3}) \notag,
\end{align}
combining both second-level wavefunctions with $(n-4)$ matrix elements $\Omega$, which diagrammatically amounts to\footnotetext{At first glimpse, their blocks contain slightly different coefficients. However, one can verify that they are actually the same by comparing the respective generating functions. Note that the coefficients $\tau_{kl}$ can be written in terms of a terminating ${}_3F_2$-hypergeometric function with unit argument.}
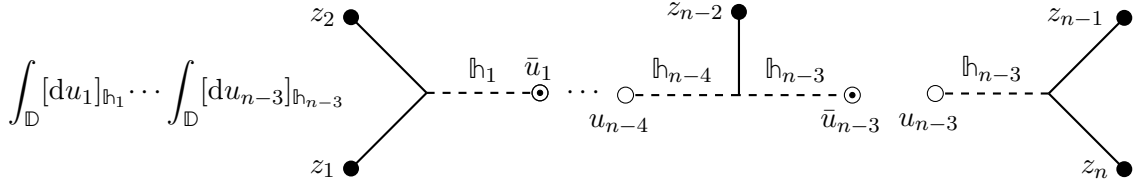
\begin{align*}
            \int_{\mathbb{D}}[\mathrm{d} u_1]_{\hbb_1}
            \dots \int_{\mathbb{D}}[\mathrm{d} u_{n-3}]_{\hbb_{n-3}} 
            \mkern-22mu \! \!
\def\tkzscl{0.5}
\begin{tikzpicture}[baseline={([yshift=-.5ex]current bounding box.center)},vertex/.style={anchor=base,
    circle,fill=black!25,minimum size=18pt,inner sep=2pt},scale=\tkzscl]
        \coordinate[label=left:$z_2$\,] (z_1) at (-2,2);
        \coordinate[label=left:$z_1$\,] (z_2) at (-2,-2);
        \coordinate[label=above:\,$\Bar{u}_1\phantom{|}$] (u_1) at (3,0);
        \draw[thick] (z_1) -- (0,0);
        \draw[thick] (z_2) -- (0,0);
        \draw[thick, dashed] (0,0) -- node[above] {$\mathbb{h}_1$} (u_1);
        \fill (z_1) circle (6pt);
        \fill (z_2) circle (6pt);
        \filldraw [fill=white] (u_1) circle (6pt);
        \filldraw (u_1) circle (2pt);
        \draw (u_1) circle (6pt);
\end{tikzpicture}
 \!\!\cdots \mkern-10mu \! \!
\def\tkzscl{0.5}
\begin{tikzpicture}[baseline={([yshift=-.5ex]current bounding box.center)},vertex/.style={anchor=base,
    circle,fill=black!25,minimum size=18pt,inner sep=2pt},scale=\tkzscl]
        \coordinate[label=left:$z_{n-2}$\,] (z_1) at (0,2.2);
        \coordinate[label=below:$u_{n-4}\phantom{|}$\,] (u_1) at (-3,0);
        \coordinate[label=below:\,$\bar{u}_{n-3}\phantom{|}$] (u_2) at (3,0);
        \draw[thick] (z_1) -- (0,0);
        \draw[thick, dashed] (u_2) --node[above] {$\mathbb{h}_{n-3}$} (0,0);
        \draw[thick, dashed] (0,0) -- node[above] {$\mathbb{h}_{n-4}$} (u_1);
        \fill (z_1) circle (6pt);
        \fill[white] (u_1) circle (6pt);
        \draw (u_1) circle (6pt);
        \fill[white] (u_2) circle (6pt);
        \filldraw (u_2) circle (2pt);
        \draw (u_2) circle (6pt);
        \draw[draw=none] (0,0) -- (0,-2.6);
\end{tikzpicture}
 \!\!
\def\tkzscl{0.5}
\begin{tikzpicture}[baseline={([yshift=-.5ex]current bounding box.center)},vertex/.style={anchor=base,
    circle,fill=black!25,minimum size=18pt,inner sep=2pt},scale=\tkzscl]
        \coordinate[label=left:$z_{n-1}$\,\,] (z_1) at (2,2);
        \coordinate[label=left:$z_n$\,] (z_2) at (2,-2);
        \coordinate[label=below:$u_{n-3}\phantom{|}$\,] (u_1) at (-3,0);
        \draw[thick] (z_1) -- (0,0);
        \draw[thick] (z_2) -- (0,0);
        \draw[thick, dashed] (0,0) -- node[above] {$\mathbb{h}_{n-3}$} (u_1);
        \fill (z_1) circle (6pt);
        \fill (z_2) circle (6pt);
        \fill[white] (u_1) circle (6pt);
        \draw (u_1) circle (6pt);
\end{tikzpicture}
\end{align*}

In principle, one can directly compute 
\eqref{eq:2d_npt_block_WFlevel} in the same way as the five-point block before.
It is more interesting in this context, however, to show-case the modular nature of the oscillator construction: given an $(n-1)$-point block in the comb channel, one can always compute the corresponding $n$-point block by using the formal relation \eqref{eq:psi_Omega_correspondence}. 
Applying this to the last $\Omega$ in the integral expression for the $n$-point block, we rediscover the $(n-1)$-point block, which can be seen diagrammatically as follows: 
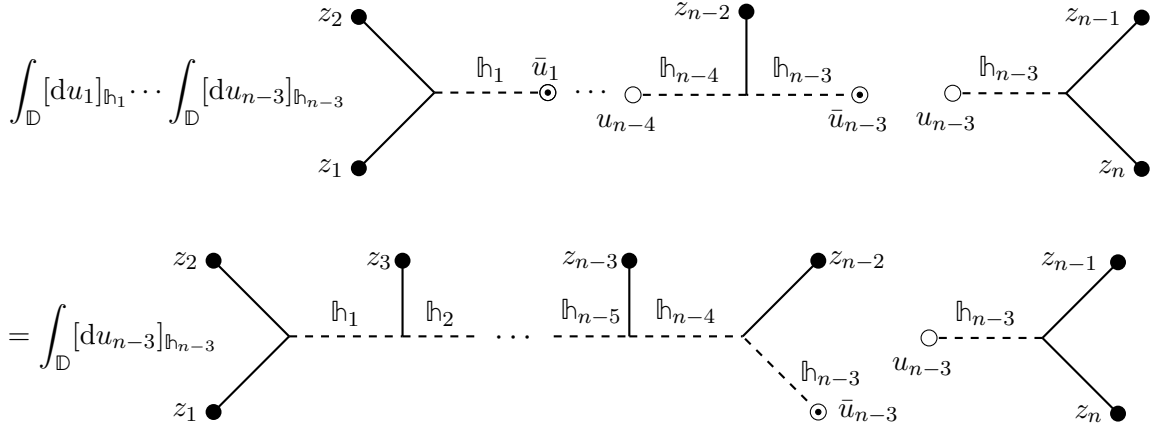
\begin{multline*}
    \int_{\mathbb{D}}[\mathrm{d} u_1]_{\hbb_1}
            \dots \int_{\mathbb{D}}[\mathrm{d} u_{n-3}]_{\hbb_{n-3}}  
            \mkern-18mu \! \!
\def\tkzscl{0.5}
\begin{tikzpicture}[baseline={([yshift=-.5ex]current bounding box.center)},vertex/.style={anchor=base,
    circle,fill=black!25,minimum size=18pt,inner sep=2pt},scale=\tkzscl]
        \coordinate[label=left:$z_2$\,] (z_1) at (-2,2);
        \coordinate[label=left:$z_1$\,] (z_2) at (-2,-2);
        \coordinate[label=above:\,$\Bar{u}_1\phantom{|}$] (u_1) at (3,0);
        \draw[thick] (z_1) -- (0,0);
        \draw[thick] (z_2) -- (0,0);
        \draw[thick, dashed] (0,0) -- node[above] {$\mathbb{h}_1$} (u_1);
        \fill (z_1) circle (6pt);
        \fill (z_2) circle (6pt);
        \filldraw [fill=white] (u_1) circle (6pt);
        \filldraw (u_1) circle (2pt);
        \draw (u_1) circle (6pt);
\end{tikzpicture}
 \!\!\cdots \mkern-10mu \! \!
\def\tkzscl{0.5}
\begin{tikzpicture}[baseline={([yshift=-.5ex]current bounding box.center)},vertex/.style={anchor=base,
    circle,fill=black!25,minimum size=18pt,inner sep=2pt},scale=\tkzscl]
        \coordinate[label=left:$z_{n-2}$\,] (z_1) at (0,2.2);
        \coordinate[label=below:$u_{n-4}\phantom{|}$\,] (u_1) at (-3,0);
        \coordinate[label=below:\,$\bar{u}_{n-3}\phantom{|}$] (u_2) at (3,0);
        \draw[thick] (z_1) -- (0,0);
        \draw[thick, dashed] (u_2) --node[above] {$\mathbb{h}_{n-3}$} (0,0);
        \draw[thick, dashed] (0,0) -- node[above] {$\mathbb{h}_{n-4}$} (u_1);
        \fill (z_1) circle (6pt);
        \fill[white] (u_1) circle (6pt);
        \draw (u_1) circle (6pt);
        \fill[white] (u_2) circle (6pt);
        \filldraw (u_2) circle (2pt);
        \draw (u_2) circle (6pt);
        \draw[draw=none] (0,0) -- (0,-2.6);
\end{tikzpicture}
\def\tkzscl{0.5}
\begin{tikzpicture}[baseline={([yshift=-.5ex]current bounding box.center)},vertex/.style={anchor=base,
    circle,fill=black!25,minimum size=18pt,inner sep=2pt},scale=\tkzscl]
        \coordinate[label=left:$z_{n-1}$\,\,] (z_1) at (2,2);
        \coordinate[label=left:$z_n$\,] (z_2) at (2,-2);
        \coordinate[label=below:$u_{n-3}\phantom{|}$\,] (u_1) at (-3,0);
        \draw[thick] (z_1) -- (0,0);
        \draw[thick] (z_2) -- (0,0);
        \draw[thick, dashed] (0,0) -- node[above] {$\mathbb{h}_{n-3}$} (u_1);
        \fill (z_1) circle (6pt);
        \fill (z_2) circle (6pt);
        \fill[white] (u_1) circle (6pt);
        \draw (u_1) circle (6pt);
\end{tikzpicture}
 \\[0.5cm]            
    = \int_{\mathbb{D}}[\d u_{n-3}]_{\hbb_{n-3}} 
    \mkern-28mu \! \!\begin{tikzpicture}[baseline={([yshift=-.5ex]current bounding box.center)},vertex/.style={anchor=base,
    circle,fill=black!25,minimum size=18pt,inner sep=2pt}]
        \coordinate[label=left:$z_1\,$] (z_1) at (-1,-1);
        \coordinate[label=left:$z_2\,$] (z_2) at (-1,1);
        \coordinate[label=left:$z_3$] (z_3) at (1.5,1);
        \coordinate[label=left:$z_{n-3}$] (z_n-2) at (4.5,1);
        \coordinate[label=right:$z_{n-2}$] (z_n-1) at (7,1);
        \coordinate[label=right:$\ \bar{u}_{n-3}$] (z_n) at (7,-1);
        \coordinate (u_1) at (1.5,0);
        \coordinate (u_2) at (2.5,0);
        \coordinate (u_3) at (3.5,0);
        \coordinate (u_4) at (4.5,0);
        \coordinate (u_5) at (6,0);
        \draw[thick] (z_1) -- (0,0);
        \draw[thick] (z_2) -- (0,0);
        \draw[thick] (z_3) -- (u_1);
        \draw[thick] (z_n-2) -- (u_4);
        \draw[thick] (z_n-1) -- (u_5);
        \draw[thick, dashed ] (z_n) -- node[right] {\ $\mathbb{h}_{n-3}$} (u_5);
        \draw[thick, dashed] (0,0) -- node[above] {$\mathbb{h}_1$} (u_1);
        \draw[thick, dashed] (u_1) -- node[above] {$\mathbb{h}_2$} (u_2);
        \draw[thick, dashed] (u_3) -- node[above] {$\mathbb{h}_{n-5}$} (u_4);
        \draw[thick, dashed] (u_4) -- node[above] {$\mathbb{h}_{n-4}$} (u_5);
        \fill (z_1) circle (3pt);
        \fill (z_2) circle (3pt);
        \fill (z_3) circle (3pt);
        \fill (z_n-2) circle (3pt);
        \fill (z_n-1) circle (3pt);
        \filldraw [fill=white] (z_n) circle (3pt);
        \filldraw (z_n) circle (1pt);
        \draw[draw=none](3.2,-0.25) -- node[above] {\scalebox{1.1}{$\cdots$}} (2.8,-0.25);
    \end{tikzpicture} 
    \mkern-10mu \! \!
\def\tkzscl{0.5}
\begin{tikzpicture}[baseline={([yshift=-.5ex]current bounding box.center)},vertex/.style={anchor=base,
    circle,fill=black!25,minimum size=18pt,inner sep=2pt},scale=\tkzscl]
        \coordinate[label=left:$z_{n-1}$\,\,] (z_1) at (2,2);
        \coordinate[label=left:$z_n$\,] (z_2) at (2,-2);
        \coordinate[label=below:$u_{n-3}\phantom{|}$\,] (u_1) at (-3,0);
        \draw[thick] (z_1) -- (0,0);
        \draw[thick] (z_2) -- (0,0);
        \draw[thick, dashed] (0,0) -- node[above] {$\mathbb{h}_{n-3}$} (u_1);
        \fill (z_1) circle (6pt);
        \fill (z_2) circle (6pt);
        \fill[white] (u_1) circle (6pt);
        \draw (u_1) circle (6pt);
\end{tikzpicture}
\end{multline*}
Translating this diagram into integrals, one gets
\begin{align}
    & G^{h_1,\dots,h_n}_{\mathbb{h}_1,\dots,\mathbb{h}_{n-3}}(z_1,\dots,z_n)   \notag\\
    &=\int_\mathbb{D}[\d u_1]_{\hbb_1}\,\dots \int_\mathbb{D}[\d u_{n-4}]_{\hbb_{n-4}}\int_\mathbb{D}[\d u_{n-3}]_{\hbb_{n-3}}\, \chi(z_1,z_2;\bar{u}_1)\Omega(z_3;u_1,\bar{u}_2)\dots \Omega(z_{n-3};u_{n-5},\bar{u}_{n-4})\notag\\
    &\hspace{8.6cm}\times \Omega(z_{n-2};u_{n-4},\bar{u}_{n-3}) 
    \psi(z_{n-1},z_n;u_{n-3}) \notag\\
    &=\int_\mathbb{D}[\d u_{n-3}]_{\hbb_{n-3}}\bigg(\! \int_\mathbb{D}[\d u_1]_{\hbb_1}\dots \!\int_\mathbb{D}[\d u_{n-4}]_{\hbb_{n-4}} \chi(z_1,z_2;\bar{u}_1)\Omega(z_3;u_1,\bar{u}_2)\dots\Omega(z_{n-3};u_{n-5},\bar{u}_{n-4}) \notag\\ 
    &\hspace{8.6cm}\times   \psi(z_{n-2},\bar{u}_{n-3};u_{n-4})\bigg)
    \psi(z_{n-1},z_n;u_{n-3}) \notag\\
    &=\int_\mathbb{D}[\d u_{n-3}]_{\hbb_{n-3}}\, G^{h_1,\dots,h_{n-2},\mathbb{h}_{n-3}}_{\mathbb{h}_1,\dots,\mathbb{h}_{n-4}} (z_1,\dots,z_{n-2},\bar{u}_{n-3})\, \psi(z_{n-1},z_n;u_{n-3}). \label{eq:induction_npt_comb}
\end{align}
This observation allows for an inductive computation, similar in spirit to \cite{Rosenhaus:2018zqn}, that leads to the same result. We give the details in appendix \ref{app:comb_npt} and here only state the form of the $n$-point comb block as
\begin{equation}
\label{eq:comb_npt_block}
    G^{h_1,\dots,h_n}_{\mathbb{h}_1,\dots,\mathbb{h}_{n-3}}(z_1,\dots,z_n)  = \mathcal{L}^{h_1,\dots,h_n}(z_1,\dots,z_n)\, g^{h_1,\dots,h_n}_{\mathbb{h}_1,\dots,\mathbb{h}_{n-3}}(\xi_1,\dots,\xi_{n-3}),
\end{equation}
where the leg factor
\begin{equation}
\label{eq:comb_npt_leg_factor}
    \mathcal{L}^{h_1,\dots,h_n}(z_1,\dots,z_n) = \left(\frac{z_{23}}{z_{12}z_{13}}\right)^{h_1} \left(\frac{z_{n-2,n-1}}{z_{n-2,n}z_{n-1,n}}\right)^{h_n}\, \prod\limits_{i=1}^{n-2} \left(\frac{z_{i,i+2}}{z_{i,i+1}z_{i+1,i+2}}\right)^{h_{i+1}}
\end{equation}
is universal for all blocks. The bare block
\begin{equation}
g^{h_1,\dots,h_n}_{\mathbb{h}_1,\dots,\mathbb{h}_{n-3}}(\xi_1,\dots,\xi_{n-3}) = \prod\limits_{i=1}^{n-3} \xi_i^{\mathbb{h}_i} 
F_K\left[
    \begin{matrix}
    h_{112}, \hbb_{123},\dots,\hbb_{n-4,n-3,n-2},h_{n-3,n,n-1} \\ 2\mathbb{h}_1,\dots,2\mathbb{h}_{n-3}
    \end{matrix}; \xi_1,\dots, \xi_{n-3}
    \right] \label{eq:SL2R_bare_conf_block}
\end{equation}
with $\hbb_{ijk}=\hbb_i+\hbb_j-h_k$ and $h_{ijk}=\hbb_i+h_j-h_k$ 
is a multi-variable hypergeometric function of the cross-ratios
\begin{equation}
\label{eq:comb_npt_cross_ratios}
    \xi_i = \frac{z_{i,i+1}z_{i+2,i+3}}{z_{i,i+2}z_{i+1,i+3}}\,. 
\end{equation}
This function $F_K$ defined as  
\begin{align}
    F_K&\left[\begin{matrix}
    a_1,b_1,\dots,b_{n-4},a_2 \\ c_1,\dots,c_{n-3}
    \end{matrix}; \xi_1,\dots, \xi_{n-3}\right] \notag\\
    &\equiv \sum\limits_{k_1,\dots,k_{n-3}}^\infty \frac{(a_1)_{k_1}(b_1)_{k_1+k_2}(b_2)_{k_2+k_3}\dots(b_{n-4})_{k_{n-4}+k_{n-3}}(a_2)_{k_{n-3}}}{(c_1)_{k_1}\dots(c_{n-3})_{k_{n-3}}} \frac{\xi_1^{k_1}}{k_1!}\dots \frac{\xi_{n-3}^{k_{n-3}}}{k_{n-3}!} \label{eq:SL2R_comb_function}
\end{align}
is introduced in \cite{Rosenhaus:2018zqn} as the so-called comb function.

\subsection{Conformal Blocks in Four Dimensions}
\label{sec:4ptblock_4d}
The four-point block $G_\Deltabb^{\Delta_1,\dots,\Delta_4}$ in four dimensions can be computed in the same way as in two dimensions, using the corresponding wavefunctions. Hence, also diagrammatically nothing changes except the labelling
\begin{equation*}
\def\tkzscl{0.5}
\begin{tikzpicture}[baseline={([yshift=-.5ex]current bounding box.center)},vertex/.style={anchor=base,
    circle,fill=black!25,minimum size=18pt,inner sep=2pt},scale=\tkzscl]
        \coordinate[label=left:$X_2$\,] (z_2) at (-2,2);
        \coordinate[label=left:$X_1$\,] (z_1) at (-2,-2);
        \coordinate[label=right:\,$X_3$] (z_3) at (5,2);
        \coordinate[label=right:\,$X_4$] (z_4) at (5,-2);
        \coordinate (u_1) at (3,0);
        \draw[thick] (z_1) -- (0,0);
        \draw[thick] (z_2) -- (0,0);
        \draw[thick] (z_3) -- (u_1);
        \draw[thick] (z_4) -- (u_1);
        \draw[thick, dashed] (0,0) -- node[above] {$\mathbb{\Delta}$} (u_1);
        \fill (z_1) circle (6pt);
        \fill (z_2) circle (6pt);
        \fill (z_3) circle (6pt);
        \fill (z_4) circle (6pt);
    \end{tikzpicture}
  =  \int_{\mathbb{D}_4}[\d U]_\Deltabb 
\def\tkzscl{0.5}
\begin{tikzpicture}[baseline={([yshift=-.5ex]current bounding box.center)},vertex/.style={anchor=base,
    circle,fill=black!25,minimum size=18pt,inner sep=2pt},scale=\tkzscl]
        \coordinate[label=left:$X_2$\,] (z_1) at (-2,2);
        \coordinate[label=left:$X_1$\,] (z_2) at (-2,-2);
        \coordinate[label=right:\,$U^\dagger\phantom{|}$] (u_1) at (3,0);
        \draw[thick] (z_1) -- (0,0);
        \draw[thick] (z_2) -- (0,0);
        \draw[thick, dashed] (0,0) -- node[above] {$\mathbb{\Delta}$} (u_1);
        \fill (z_1) circle (6pt);
        \fill (z_2) circle (6pt);
        \filldraw [fill=white] (u_1) circle (6pt);
        \filldraw (u_1) circle (2pt);
        \draw (u_1) circle (6pt);
\end{tikzpicture}
\def\tkzscl{0.5}
\begin{tikzpicture}[baseline={([yshift=-.5ex]current bounding box.center)},vertex/.style={anchor=base,
    circle,fill=black!25,minimum size=18pt,inner sep=2pt},scale=\tkzscl]
        \coordinate[label=right:\,$X_3$] (z_1) at (2,2);
        \coordinate[label=right:\,$X_4$] (z_2) at (2,-2);
        \coordinate[label=left:$U$\,] (u_1) at (-3,0);
        \draw[thick] (z_1) -- (0,0);
        \draw[thick] (z_2) -- (0,0);
        \draw[thick, dashed] (0,0) -- node[above] {$\mathbb{\Delta}$} (u_1);
        \fill (z_1) circle (6pt);
        \fill (z_2) circle (6pt);
        \fill[white] (u_1) circle (6pt);
        \draw (u_1) circle (6pt);
\end{tikzpicture}
\end{equation*}
More specifically, we use global symmetry to set 
\begin{equation}
    X_1 \to \infty\,, \qquad X_2 = \mathbb{1}\,,\qquad X \equiv X_3 = \begin{pmatrix}
    x+iy & 0 \\ 0 & x-iy
\end{pmatrix}\,, \qquad X_4 = 0\,,
\end{equation}
such that the second-level wavefunction $\chi$ reduces to
the following expression 
\begin{align}
\label{eq:chi_exp_4pt_4d}
    \chi(\infty, \mathbb{1};U^\dagger) = x_1^{-2\Delta_1}\sum\limits_{q_a,q_b}^{j,m} \left(\mathcal{N}^{j,m}_{\beta}\right)^2 \overline{  \phi_{q_a,q_b}^{j,m}(U)}\, \phi_{q_a,q_b}^{j,m}(\mathbb{1})\,,
\end{align}
where $x_1^{-2\Delta_1}$ is the leading order of the $x_1 \to \infty$ limit as $\Delta_1=\alpha+\gamma$, and $\beta = \frac{1}{2}(\mathbb{\Delta}+\Delta_2-\Delta_1)$. Note that even though $X_2 = \mathbb{1}$ does not lie in $\mathbb{D}_4$, one can separately check that this expansion converges. Similarly to equation \eqref{eq:chi_exp_4pt_4d}, the second-level wavefunctions $\psi$ containing the remaining two points can be expanded to 
\begin{align}
    \psi(X, 0; U) = \mathrm{det}^{-\delta}(X) \sum\limits_{q'_a,q'_b}^{j',m'} \left(\mathcal{N}^{j',m'}_{\varepsilon}\right)^2 \phi_{q'_a,q'_b}^{j',m'}(U)\, \phi_{q'_a,q'_b}^{j',m'}(X^T) \label{eq:4d_4pt_coordinate_choice}
\end{align}
with $\delta = \frac{1}{2}(\Delta_3 +\Delta_4-\mathbb{\Delta})$ and $\varepsilon = \frac{1}{2}(\mathbb{\Delta} + \Delta_3 - \Delta_4)$. Assembling these ingredients, the four-point block reads
\begin{align}
     G_\Deltabb^{\Delta_1,\dots,\Delta_4} =&\expval{\mathcal{O}_{\Delta_1}(\infty)\mathcal{O}_{\Delta_2}(\mathbb{1})\mathbb{P}_\mathbb{\Delta} \mathcal{O}_{\Delta_3}(X)\mathcal{O}_{\Delta_4}(0)} \notag\\
    =& \int_{\Disc_4} [\dd U]_{\mathbb{\Delta}}\, \chi(\infty, \mathbb{1};U^\dagger) \psi(X, 0;U) \notag\\
    =&\, \mathrm{det}^{-\delta}(X) \sum\limits_{q_a,q_b}^{j,m} \sum\limits_{q_a',q_b'}^{j',m'} \left(\mathcal{N}^{j,m}_{\beta} \mathcal{N}^{j',m'}_{\varepsilon}\right)^2\! \phi_{q_a,q_b}^{j,m}(\mathbb{1})  \phi_{q_a',q_b'}^{j',m'}(X^T) \int_{\Disc_4}\! [\dd U]_\Deltabb\, \overline{ \phi_{q_a,q_b}^{j,m}(U)} \phi_{q_a',q_b'}^{j',m'}(U) \notag\\
    =& \,\mathrm{det}^{-\delta}(X) \sum\limits_{q_a,q_b}^{j,m} \sum\limits_{q_a',q_b'}^{j',m'} \phi_{q_a,q_b}^{j,m}(\mathbb{1})\,  \phi_{q_a',q_b'}^{j',m'}(X^T)\, \frac{\left(\mathcal{N}^{j,m}_{\beta}\mathcal{N}^{j',m'}_{\varepsilon}\right)^2}{\left(\mathcal{N}^{j,m}_{\mathbb{\Delta}}\right)^2}\, \delta^{j',m',q_a',q_b'}_{j,m,q_a,q_b}\\
    \equiv& \,\mathrm{det}^{-\delta}(X) \sum\limits_{q_a,q_b}^{j,m} \left(\mathcal{N}^{j,m}_{\beta,\varepsilon;\Deltabb}\right)^2 \delta_{q_a,q_b} \mathrm{det}^m(X) \mathcal{D}^j_{q_a,q_b}(X^T)\,.\notag
\end{align}
We abbreviated the coefficients as $\left(\mathcal{N}^{j,m}_{\beta,\varepsilon;\Deltabb}\right)^2$ and used that $\mathcal{D}^j_{q_a,q_b}(\mathbb{1}) = \delta_{q_a,q_b}$. Now, explicitly implementing the choice of a diagonal matrix $X$ in \eqref{eq:4d_4pt_coordinate_choice} simplifies the Wigner-$\mathcal{D}$ matrix. Moving to complex coordinates \mbox{$z = x+iy$} and $\bar{z} = x-iy $, we obtain 
\begin{align}
\begin{split}
    G_\Deltabb^{\Delta_1,\dots,\Delta_4} &=  (z\bar{z})^{\frac{1}{2}(\Deltabb-\Delta_3-\Delta_4)} \sum\limits_{m=0}^\infty \sum\limits_{j\in\mathbb{N}/2} \sum\limits_{q=-j}^j \left(\mathcal{N}^{j,m}_{\beta,\varepsilon;\Deltabb}\right)^2\! z^{m+j+q} \bar{z}^{m+j-q} \\
    &=  (z\bar{z})^{\frac{1}{2}(\Deltabb-\Delta_3-\Delta_4)} \sum\limits_{m=0}^\infty \sum\limits_{j\in\mathbb{N}/2}  \left(\mathcal{N}^{j,m}_{\beta,\varepsilon;\Deltabb}\right)^2\! z^m \bar{z}^m\sum\limits_{k=0}^{2j} z^{k} \bar{z}^{2j-k} \\
    &=  \frac{(z\bar{z})^{\frac{1}{2}(\Deltabb-\Delta_3-\Delta_4)}}{z-\bar{z}} \sum\limits_{m=0}^\infty \sum\limits_{j\in\mathbb{N}/2}  \left(\mathcal{N}^{j,m}_{\beta,\varepsilon;\Deltabb}\right)^2\! z^m \bar{z}^m (z^{2j+1} - \bar{z}^{2j+1}) \,.
\end{split}
\end{align}
In the last step, we applied the geometric series. The prefactor of the above result contains a residue of the leg factor that is $(z\bar{z})^{\frac{1}{2}(-\Delta_3-\Delta_4)}$. Stripping that off, we obtain the bare block
\begin{align}
\label{eq:result_4pt_4d}    g_\Deltabb^{\Delta_1,\dots,\Delta_4} &= \frac{(z\bar{z})^{\frac{\Deltabb}{2}}}{z-\bar{z}} \sum\limits_{m=0}^\infty \sum\limits_{j\in\mathbb{N}/2}  \left(\mathcal{N}^{j,m}_{\beta,\varepsilon;\Deltabb}\right)^2 z^m \bar{z}^m (z^{2j+1} - \bar{z}^{2j+1})\,.
\end{align}
We show in appendix \ref{app:correspondence_4p} that this indeed matches the result in \cite{Dolan:2000ut,Dolan:2003hv}, as the following equality holds: 
\begin{align}
    &\sum\limits_{m=0}^\infty \sum\limits_{j\in\mathbb{N}/2}  \left(\mathcal{N}^{j,m}_{\beta,\varepsilon;\Deltabb}\right)^2 z^m \bar{z}^m \left(z^{2j+1} - \bar{z}^{2j+1}\right) \label{eq:4ptBlock_series_comparisson}\\
    &\quad=  z\, {}_2F_1\! \left( \begin{matrix}
        \beta, \varepsilon \\ \Deltabb
    \end{matrix}; z \right) {}_2F_1\! \left( \begin{matrix}
        \beta - 1, \varepsilon - 1 \\ \Deltabb -2
    \end{matrix}; \bar{z} \right)
    -\bar{z}\, {}_2F_1\! \left( \begin{matrix}
        \beta, \varepsilon \\ \Deltabb
    \end{matrix}; \bar{z} \right) {}_2F_1\! \left( \begin{matrix}
        \beta - 1, \varepsilon - 1 \\ \Deltabb -2
    \end{matrix}; z \right)\,. \notag
\end{align}
Note that both expressions are anti-symmetric in $z$ and $\bar{z}$, since the overall block must be symmetric.

Throughout the calculation we used the expansion of the wavefunctions in the basis through \eqref{eq:expansion_kernel} which technically only holds for $\Deltabb \geq 2$. However, as the conformal block is analytic in $\Deltabb$, we can always analytically continue in $\Deltabb$ to the general solution. Another option is to directly check if the solution fulfils the Casimir equations for general weights -- which it does.

Another issue with the expansion \eqref{eq:expansion_kernel} that we have not discussed so far, is that it was originally given for the kernel, which only depends on oscillator variables. However, we also used it for wavefunctions $\psi$ and $\chi$ that both depend on coordinates that a priori do not have to lie in the ball $\mathbb{D}_4$ at all. But if we consider the conformal blocks which we are computing to arise from an operator product expansion, we already have restrictions on their coordinates: in the case of the four-point block, we need to be able to separate e.g.  $\mathcal{O}_{\Delta_1}(X_1)$ and $\mathcal{O}_{\Delta_2}(X_2)$ from $\mathcal{O}_{\Delta_3}(X_3)$ and $\mathcal{O}_{\Delta_4}(X_4)$ through a topological hypersurface such that the enclosed space has no further operator insertions and can be conformally mapped to an open ball. If we assume this for $X_1$ to $X_4$, all the coordinates that the wavefunctions depend on lie inside an open ball and they can be expanded in terms of the basis.

\subsubsection{Towards Higher-point Conformal Blocks in Four Dimensions}

Finally, let us consider a special case of the $n$-point block in the comb channel as an example illustrating how the oscillator formalism can be applied for the computation of higher-point blocks in four dimension. Schematically, the construction works precisely as demonstrated in the two-dimensional case: starting with the five-point block, it is obtained by glueing three second-level oscillator wavefunctions $\chi$, $\Omega$ and $\psi$ via two integrations over $\Disc_4$ as follows 
\begin{multline*}
    \def\tkzscl{0.5}
\begin{tikzpicture}[baseline={([yshift=-.5ex]current bounding box.center)},vertex/.style={anchor=base,
    circle,fill=black!25,minimum size=18pt,inner sep=2pt},scale=\tkzscl]
        \coordinate[label=left:$X_2$\,] (z_2) at (-2,2);
        \coordinate[label=left:$X_1$\,] (z_1) at (-2,-2);
        \coordinate[label=above:$X_3$] (z_3) at (3,2.83);
        \coordinate[label=right:\,$X_4$] (z_4) at (8,2);
        \coordinate[label=right:\,$X_5$] (z_5) at (8,-2);
        \coordinate (u_1) at (3,0);
        \coordinate (u_2) at (6,0);
        \draw[thick] (z_1) -- (0,0);
        \draw[thick] (z_2) -- (0,0);
        \draw[thick] (z_3) -- (u_1);
        \draw[thick] (z_4) -- (u_2);
        \draw[thick] (z_5) -- (u_2);
        \draw[thick, dashed] (0,0) -- node[above] {$\mathbb{\Delta}_1$} (u_1);
        \draw[thick, dashed] (u_1) -- node[above] {$\mathbb{\Delta}_2$} (u_2);
        \fill (z_1) circle (6pt);
        \fill (z_2) circle (6pt);
        \fill (z_3) circle (6pt);
        \fill (z_4) circle (6pt);
        \fill (z_5) circle (6pt);
\end{tikzpicture}  \\
    =\!  \int_{\mathbb{D}_4}\![\d U_1]_{\Deltabb_1}\!\!\int_{\mathbb{D}_4}\![\d U_2]_{\Deltabb_2} \hspace{-.6cm}
\def\tkzscl{0.5}
\begin{tikzpicture}[baseline={([yshift=-.5ex]current bounding box.center)},vertex/.style={anchor=base,
    circle,fill=black!25,minimum size=18pt,inner sep=2pt},scale=\tkzscl]
        \coordinate[label=left:$X_2$\,] (z_1) at (-2,2);
        \coordinate[label=left:$X_1$\,] (z_2) at (-2,-2);
        \coordinate[label=right:\,$U_1^\dagger\phantom{|}$] (u_1) at (2.5,0);
        \draw[thick] (z_1) -- (0,0);
        \draw[thick] (z_2) -- (0,0);
        \draw[thick, dashed] (0,0) -- node[above] {$\mathbb{\Delta}_1$} (u_1);
        \fill (z_1) circle (6pt);
        \fill (z_2) circle (6pt);
        \filldraw [fill=white] (u_1) circle (6pt);
        \filldraw (u_1) circle (2pt);
        \draw (u_1) circle (6pt);
\end{tikzpicture}
 \!
\def\tkzscl{0.5}
\begin{tikzpicture}[baseline={([yshift=-.5ex]current bounding box.center)},vertex/.style={anchor=base,
    circle,fill=black!25,minimum size=18pt,inner sep=2pt},scale=\tkzscl]
        \coordinate[label=left:\,$U_1\phantom{|}$] (u_1) at (-2.5,0);
        \coordinate[label=right:\,$U_2^\dagger\phantom{|}$](u_2) at (2.5,0);
        \coordinate[label=above:$X_3$]  (z_1) at (0,2.83);
        \draw[thick] (z_1) -- (0,0);
        \draw[thick, dashed] (0,0) -- node[above] {$\mathbb{\Delta}_1$} (u_1);
        \draw[thick, dashed] (0,0) -- node[above] {$\mathbb{\Delta}_2$} (u_2);
        \fill (z_1) circle (6pt);
        \fill[white] (u_1) circle (6pt);
        \draw (u_1) circle (6pt);
        \filldraw [fill=white] (u_2) circle (6pt);
        \filldraw (u_2) circle (2pt);
        \draw (u_2) circle (6pt);
        \draw[draw=none] (0,0) -- (0,-3.9);
\end{tikzpicture}
 \!
\def\tkzscl{0.5}
\begin{tikzpicture}[baseline={([yshift=-.5ex]current bounding box.center)},vertex/.style={anchor=base,
    circle,fill=black!25,minimum size=18pt,inner sep=2pt},scale=\tkzscl]
        \coordinate[label=right:\,$X_4$] (z_1) at (2,2);
        \coordinate[label=right:\,$X_5$] (z_2) at (2,-2);
        \coordinate[label=left:$U_2$\,] (u_1) at (-2.5,0);
        \draw[thick] (z_1) -- (0,0);
        \draw[thick] (z_2) -- (0,0);
        \draw[thick, dashed] (0,0) -- node[above] {$\mathbb{\Delta}_2$} (u_1);
        \fill (z_1) circle (6pt);
        \fill (z_2) circle (6pt);
        \fill[white] (u_1) circle (6pt);
        \draw (u_1) circle (6pt);
\end{tikzpicture}
\end{multline*}
We then use equations \eqref{eq:sol_chi_2ndLvl_V1}, \eqref{eq:sol_psi_2ndLvl_V1} and \eqref{eq:sol_Omega_2ndLvl_V1} such that the five-point block reads
\begin{align}
\begin{split}
    &\expval{\mathcal{O}_1(\infty)\mathcal{O}_2(\mathbb{1})\mathbb{P}_{\Deltabb_1}{\mathcal{O}_3(X_3)\mathbb{P}_{\Deltabb_2}\mathcal{O}_4(X_4)\mathcal{O}_5(0)}} \\ 
    &= \int_{\mathbb{D}_4}[\d U_1]_{\Deltabb_1}\int_{\mathbb{D}_4}[\d U_2]_{\Deltabb_2} \chi(\infty,\mathbb{1};U_1^\dagger) \Omega(X_3;U_1,U_2^\dagger) \psi(X_4,0;U_2) \\
    &= \mathrm{det}^{-\tilde{\gamma}}(X_3)\, \mathrm{det}^{-\gamma'}(X_4) \sum\limits_{(1),\dots,(5)} \phi^{(1)}(\mathbb{1})\phi^{(2)}(X_3^{\mathrm{T}}) \phi^{(4)}(X_3^{-1}) \phi^{(5)}(X_4^{\mathrm{T}}) \\
    &\quad \times \left( \mathcal{N}^{(1)}_{\beta} \mathcal{N}^{(2)}_{\tilde{\alpha}} \mathcal{N}^{(3)}_{\tilde{\beta}} \mathcal{N}^{(4)}_{\tilde{\gamma}} \mathcal{N}^{(5)}_{\alpha'}\right)^2 \int_{\mathbb{D}_4}[\d U_1]_{\Deltabb_1} \overline{\phi^{(1)}(U_1)} \phi^{(2)}(U_1) \phi^{(3)}(U_1) \\
    &\quad \times \int_{\mathbb{D}_4}[\d U_2]_{\Deltabb_2} \overline{\phi^{(3)}(U_2)\phi^{(4)}(U_2)} \phi^{(5)}(U_2)\,,
\end{split}
\end{align}
where we applied the expansion in base functions as discussed above and used the short-hand notation $(k) = (j_k,m_k,q_{a,k},q_{b,k})$ for $k=1,\dots,5$. We recognise the $\mathbb{D}_4$-integrals as the inner products $\braket{\phi^{(1)}}{\phi^{(2)}\cdot \phi^{(3)}}$ and $\braket{\phi^{(3)}\cdot\phi^{(4)}}{\phi^{(5)}}$, respectively. Instead of explicitly evaluating these integrals, we now use the expansion 
\begin{equation}
   \phi^{(2)}(U) \phi^{(3)}(U) = \sum_{(1)} \left(\mathcal{N}^{(1)}_{\Deltabb_1}\right)^2 \braket{\phi^{(1)}}{\phi^{(2)}\cdot\phi^{(3)}}_{\Deltabb_1} \phi^{(1)}(U)
\end{equation}
which follows from orthogonality. 
In the special case of $\beta = \Deltabb_1$ and $\alpha' = \Deltabb_2$, i.\,e. $\Deltabb_1 = \Delta_2-\Delta_1$ and $\Deltabb_2 = \Delta_4-\Delta_5$, we can resum the expansion to arrive at the following compact expression for the five-point block:
\begin{align}
    &\expval{\mathcal{O}_1(\infty)\mathcal{O}_2(\mathbb{1})\mathbb{P}_{\Deltabb_1}{\mathcal{O}_3(X_3)\mathbb{P}_{\Deltabb_2}\mathcal{O}_4(X_4)\mathcal{O}_5(0)}} \notag \\
    &= {\det}^{-\tilde{\gamma}}(X_3)\, {\det}^{-\gamma'}(X_4) \sum_{(2),(3),(4)} \left(\mathcal{N}^{(2)}_{\tilde{\alpha}} \mathcal{N}^{(3)}_{\tilde{\beta}}\mathcal{N}^{(4)}_{\tilde{\gamma}}\right)^2 \phi^{(2)}(X_3^T)\phi^{(4)}(X^{-1}_3) \phi^{(2)}(\mathbb{1})\phi^{(3)}(\mathbb{1}) \notag\\ &\quad\times \phi^{(3)}(X_4^{T}) \phi^{(4)}(X_4^{T}) \notag \\
    &= {\det}^{-\gamma'}(X_4)\, {\det}^{-\tilde{\alpha}}(\mathbb{1}-X_3)\, {\det}^{-\tilde{\beta}}(\mathbb{1}-X_4)\,{\det}^{-\tilde{\gamma}}(X_3-X_4)\,. 
\end{align}
Written in coordinates that gives
\begin{multline}
    \expval{\mathcal{O}_1(\infty)\mathcal{O}_2(\mathbb{1})\mathbb{P}_{\Deltabb_1}{\mathcal{O}_3(X_3)\mathbb{P}_{\Deltabb_2}\mathcal{O}_4(X_4)\mathcal{O}_5(0)}} \\
    =\abs{x_{23}}^{\Delta_1-\Delta_2-\Delta_3+\Delta_4-\Delta_5} \abs{x_{24}}^{\Delta_1-\Delta_2+\Delta_3-\Delta_4+\Delta_5} \abs{x_{34}}^{-\Delta_1+\Delta_2-\Delta_3-\Delta_4+\Delta_5} \abs{x_{4}}^{-2\Delta_5}
\end{multline}
with $x_2=(1,0,0,0)$. Indeed, this result is consistent with the one obtained in \cite{Rosenhaus:2018zqn}. Note, however, that the simplicity of this special case is immediately clear from the oscillator method while remaining far less apparent in the construction from the shadow formalism.  
The generalisation of this special case to the $n$-point conformal block being an analogous product of kernels can be seen from the $\Omega$-wavefunction in equation \eqref{eq:sol_Omega_2ndLvl_V1}: choosing inner weights such that for every additional $\Omega$-wavefunction the exponent $\tilde{\beta}$ vanishes, reduces the number of oscillator variables such that resummation after integration is possible in the same way as for the five-point block presented above.\footnote{In the two dimensional case this can be checked also on the level of coefficients; the analogous choice of weights reduces the $n$-point comb block to a product of kernels, when written as a product of $\tau$-coefficients in equation \eqref{eq:tau_coeffs}.} This allows us to write the $n$-point comb block for general external and one particular set of internal weights simply as a product of coordinate differences 
\begin{equation}
    \expval{\mathcal{O}_1(\infty)\mathcal{O}_2(\mathbb{1})\mathbb{P}_{\Deltabb_1}{\mathcal{O}_3(X_3)\mathbb{P}_{\Deltabb_2}\dots\mathbb{P}_{\Deltabb_{n-3}}\mathcal{O}_{n-1}(X_{n-1})\mathcal{O}_n(0)}} = \prod_{1<i<j<n}\abs{x_{ij}}^{\kappa_{ij}}
\end{equation}
whose exponents $\kappa_{ij}$ are linear combinations of the external weights. Note that this special case is not apparent from for example the result given in \cite{Fortin:2019zkm}. It is worth mentioning that we can also evaluate the $n$-point comb channel block for general inner weights, however, the resulting expression is hard to resum. We will report on that in a future publication.

\section{Summary and Outlook} 
We have reviewed the oscillator formalism for the two-dimensional global conformal algebra and put it to use by computing the $n$-point comb block. We highlighted the relative simplicity of the approach compared to other derivations. Similar diagrammatic representations of blocks as in for example \cite{Fortin:2020zxw,Hoback:2020pgj,Fortin:2022grf} are conceptually different and vary in complexity. The provided framework allows to write down the series expansion of a block after then trivial integrations, with no further steps. This is for example in contrast to the shadow formalism used in \cite{Rosenhaus:2018zqn}, where one has to find several convenient coordinate transformations. The wavefunctions serve as generating functions for the coefficients appearing in the series expansion of the conformal block, while simultaneously providing a geometric intuition for generalisations. 

We have not discussed conformal blocks in channels of different topologies on which there has been some focus recently \cite{Cho:2017oxl,Fortin:2020bfq,Fortin:2020yjz} and which are especially interesting from a holographic perspective \cite{Haehl2020}. For many of these channels that have not yet been computed the oscillator formalism may prove useful because of its toolbox nature.
We then generalised the construction to the four-dimensional case, where we restricted ourselves to the scalar four-point block with scalar exchange as a proof of concept to compare the approach in two and four dimensions. The higher-point blocks are computable with the same approach, but one has to solve integrals over products of three and more basis functions; this will be expanded on in future work. As an outlook, we have discussed a special case of the $n$-point block in the comb channel that allows for resummation.

These integrals over the general products of the basis functions then also can be used to compute thermal conformal blocks, by allowing for self-glueing of the $\Omega$-wavefunction. In the case of the one-point block in four dimensions, the corresponding diagram is 
\begin{align}
    \expval{\mathbb{P}_{\Deltabb}\mathcal{O}_\Delta}_\beta \quad = \quad
    \begin{tikzpicture}[baseline={([yshift=-0.65ex]current bounding box.center)},vertex/.style={anchor=base,
    circle,fill=black!25,minimum size=18pt,inner sep=2pt}]
        \coordinate[] (u) at (5,1);
        \coordinate[] (belowu) at (5.25,1);
        \coordinate[] (ubar) at (5,-1);
        \coordinate[] (aboveubar) at (5.25,-1);
        \coordinate[] (w_1) at (2.5,0);
        \draw[thick,dashed] (u) -- node[above left] {$\Deltabb$} (4,0);
        \draw[thick,dashed] (ubar) -- node[below left] {$\Deltabb$} (4,0);
        \draw[thick] (4,0) --node[above] {$\Delta$} (w_1);
        \fill (w_1) circle (3pt);
        \draw (u) circle (3pt);
        \draw[{Latex[round]}-{Latex[round]},solid] (aboveubar) to[out=-10,in=10, looseness=1.5] (belowu);
        \fill[white] (u) circle (3pt);
        \fill[white] (ubar) circle (3pt);
        \filldraw (ubar) circle (1pt);
        \draw (ubar) circle (3pt);
    \end{tikzpicture}
    = \quad 
    \begin{tikzpicture}[baseline={([yshift=-.65ex]current bounding box.center)},vertex/.style={anchor=base,
    circle,fill=black!25,minimum size=18pt,inner sep=2pt}]
    \coordinate[] (w_1) at (2.5,0);
    \fill (w_1) circle (3pt);
    \draw[thick] (3.95,0) --node[above] {$\Delta$} (w_1);
     \draw[thick,dashed] (5,0) circle (30pt);
     \coordinate[label=above right:$\Deltabb$] (h) at (6.1,0);
    \end{tikzpicture}
\end{align}
where we denoted the glueing operation by a double arrow. This enables the computation of objects like projection channel blocks in both two and four dimensions. It is then straightforward to reproduce for example the $n$-point thermal block in two dimensions, given in \cite{Alkalaev:2022kal}, as well as the one-point thermal block in four dimensions obtained in \cite{Gobeil:2018fzy}. More generally, this approach allows for a direct series expansion of $n$-point thermal blocks.

Other ongoing work in four dimensions includes the generalisation to spinning scalar conformal blocks \cite{Costa_2011} as well as blocks containing external primary operators with spin.
An exciting prospect to explore is how the vertex operators introduced in \cite{Buric:2020dyz,Buric:2021ywo, Buric:2021ttm, Buric:2021kgy} can be accounted for in this framework.

Lastly, it would be very compelling to understand the construction for non-highest weight representations, such as the principal series of $\mathfrak{sl}(2,\mathbb{C})$ and the induced representation of $\mathfrak{bms}_3$ or $\mathfrak{bms}_4$, which are relevant in the context of celestial and Carrollian CFTs, and explore how this connects to the shadow formalism (\cite{Pasterski:2016qvg, Pasterski:2017kqt, Atanasov:2021cje}, for more recent work see \cite{Liu:2024lbs, Pacifico:2025emk}). The oscillator formalism for the highest-weight representations of $\mathfrak{bms}_3$ has been introduced in \cite{Ammon:2020wem}.

\section*{Acknowledgements}
The authors thank Volker Schomerus, Christoph Sieling and José Diogo Simão for helpful discussions, as well as Michel Pannier for his help regarding computations of generating functions. Furthermore, the authors are grateful to Jonah Baerman, Christoph Hinkel and Nathaniel Riemer for fruitful discussions and related work on the topic. The authors thank ESI Vienna for providing an environment for productive discussions during the ESI Workshop on Carrollian Physics and Holography 2024. 

The work of JH, TH and KW is funded by the \emph{Deutsche Forschungsgemeinschaft (DFG)} under Grant No.\,406116891 within the Research Training Group RTG\,2522/1. JH is funded by a \emph{Landesgraduiertenstipendium} of the federal state of Thuringia.

\begin{appendices}
\addtocontents{toc}{\protect\setcounter{tocdepth}{2}}
\section{Unitarity Bounds}
We check whether our representation of $\mathfrak{su}(2,2)$ on $\mathcal{H}L^2(\mathbb{D}_4)$ indeed constitutes a unitary representation or more precisely, we check whether the corresponding representation of the Euclidean conformal algebra satisfies reflection positivity. To do so, we first provide how the generators act on the basis of $\mathcal{H}L^2(\mathbb{D}_4)$.

\subsection{Action of Generators on Basis}
\label{app:action_generators_basis}
For the following part, it turns out that it is more convenient to use the normalised basis
\begin{equation}
    \varphi_{q_a,q_b}^{j,m} = \mathcal{N}^{j,m}_\mathbb{\Delta} \phi_{q_a,q_b}^{j,m}\,, \qquad ( \varphi_{q_a,q_b}^{j,m} , \varphi_{q_a',q_b'}^{j',m'}) = \delta_{j,j'}\delta_{m,m'}\delta_{q_a,q_a'}\delta_{q_b,q_b'}\,.
\end{equation}
A tedious calculation shows how the $\mathfrak{su}(2,2)$ generators act on the basis functions $\varphi_{q_a,q_b}^{j,m}$ of the Hilbert space. In fact, said polynomials are eigenfunctions of the dilation operator, i.\,e. 
\begin{equation}
    \mathfrak{D} \varphi_{q_a,q_b}^{j,m} = (2j+2m+\mathbb{\Delta})\varphi_{q_a,q_b}^{j,m}\,.
\end{equation}
The generators $\mathfrak{P}^\mu$ of spacetime translations act as
\begin{subequations}
\begin{align*}\label{eq:P_on_Phi}
    \mathfrak{P}^0 \varphi_{q_a,q_b}^{j,m} &= C_{q_a,q_b}^{j,m+2j+1} \varphi_{q_a-\nicefrac{1}{2},q_b-\nicefrac{1}{2}}^{j-\nicefrac{1}{2},m} + C_{-q_a+\nicefrac{1}{2},-q_b+\nicefrac{1}{2}}^{j+\nicefrac{1}{2},m} \varphi_{q_a-\nicefrac{1}{2},q_b-\nicefrac{1}{2}}^{j+\nicefrac{1}{2},m-1} \numberthis\\
    &\quad + C_{-q_a,-q_b}^{j,m+2j+1} \varphi_{q_a+\nicefrac{1}{2},q_b+\nicefrac{1}{2}}^{j-\nicefrac{1}{2},m} + C_{q_a+\nicefrac{1}{2},q_b+\nicefrac{1}{2}}^{j+\nicefrac{1}{2},m} \varphi_{q_a+\nicefrac{1}{2},q_b+\nicefrac{1}{2}}^{j+\nicefrac{1}{2},m-1} \\
    \mathfrak{P}^1 \varphi_{q_a,q_b}^{j,m} &= -iC_{-q_a,q_b}^{j,m+2j+1} \varphi_{q_a+\nicefrac{1}{2},q_b-\nicefrac{1}{2}}^{j-\nicefrac{1}{2},m} +i C_{q_a+\nicefrac{1}{2},-q_b+\nicefrac{1}{2}}^{j+\nicefrac{1}{2},m} \varphi_{q_a+\nicefrac{1}{2},q_b-\nicefrac{1}{2}}^{j+\nicefrac{1}{2},m-1} \numberthis\\
    &\quad -i C_{q_a,-q_b}^{j,m+2j+1} \varphi_{q_a-\nicefrac{1}{2},q_b+\nicefrac{1}{2}}^{j-\nicefrac{1}{2},m} +i C_{-q_a+\nicefrac{1}{2},q_b+\nicefrac{1}{2}}^{j+\nicefrac{1}{2},m} \varphi_{q_a-\nicefrac{1}{2},q_b+\nicefrac{1}{2}}^{j+\nicefrac{1}{2},m-1} \\
    \mathfrak{P}^2 \varphi_{q_a,q_b}^{j,m} &= C_{-q_a,q_b}^{j,m+2j+1} \varphi_{q_a+\nicefrac{1}{2},q_b-\nicefrac{1}{2}}^{j-\nicefrac{1}{2},m} - C_{q_a+\nicefrac{1}{2},-q_b+\nicefrac{1}{2}}^{j+\nicefrac{1}{2},m} \varphi_{q_a+\nicefrac{1}{2},q_b-\nicefrac{1}{2}}^{j+\nicefrac{1}{2},m-1} \numberthis\\
    &\quad - C_{q_a,-q_b}^{j,m+2j+1} \varphi_{q_a-\nicefrac{1}{2},q_b+\nicefrac{1}{2}}^{j-\nicefrac{1}{2},m} + C_{-q_a+\nicefrac{1}{2},q_b+\nicefrac{1}{2}}^{j+\nicefrac{1}{2},m} \varphi_{q_a-\nicefrac{1}{2},q_b+\nicefrac{1}{2}}^{j+\nicefrac{1}{2},m-1} \\
    \mathfrak{P}^3 \varphi_{q_a,q_b}^{j,m} &= -iC_{q_a,q_b}^{j,m+2j+1} \varphi_{q_a-\nicefrac{1}{2},q_b-\nicefrac{1}{2}}^{j-\nicefrac{1}{2},m} -i C_{-q_a+\nicefrac{1}{2},-q_b+\nicefrac{1}{2}}^{j+\nicefrac{1}{2},m} \varphi_{q_a-\nicefrac{1}{2},q_b-\nicefrac{1}{2}}^{j+\nicefrac{1}{2},m-1} \numberthis\\
    &\quad +i C_{-q_a,-q_b}^{j,m+2j+1} \varphi_{q_a+\nicefrac{1}{2},q_b+\nicefrac{1}{2}}^{j-\nicefrac{1}{2},m} +i C_{q_a+\nicefrac{1}{2},q_b+\nicefrac{1}{2}}^{j+\nicefrac{1}{2},m} \varphi_{q_a+\nicefrac{1}{2},q_b+\nicefrac{1}{2}}^{j+\nicefrac{1}{2},m-1} 
\end{align*}
\end{subequations}
The generators $\mathfrak{K}^\mu$ of special conformal transformations act as\footnote{Note that the notation has been clarified compared to \cite{Calixto:2014sfa}. We changed for example the coefficient $C^{j+1,m+2j+1}_{-q_a + \nicefrac{1}{2},-q_b + \nicefrac{1}{2}}$ to $C^{j+1,m+2j+2}_{-q_a + \nicefrac{1}{2},-q_b + \nicefrac{1}{2}}$.}
\begin{subequations}\label{eq:K_on_Phi}
\begin{align*} 
    \mathfrak{K}^0 \varphi_{q_a,q_b}^{j,m} &= -C_{q_a,q_b}^{j,m+1} \varphi_{q_a-\nicefrac{1}{2},q_b-\nicefrac{1}{2}}^{j-\nicefrac{1}{2},m+1} - C_{-q_a,-q_b}^{j,m+1} \varphi_{q_a+\nicefrac{1}{2},q_b+\nicefrac{1}{2}}^{j-\nicefrac{1}{2},m+1} \numberthis\\
    &\quad - C_{-q_a+\nicefrac{1}{2},-q_b+\nicefrac{1}{2}}^{j+\nicefrac{1}{2},m+2j+2} \varphi_{q_a-\nicefrac{1}{2},q_b-\nicefrac{1}{2}}^{j+\nicefrac{1}{2},m} - C_{q_a+\nicefrac{1}{2},q_b+\nicefrac{1}{2}}^{j+\nicefrac{1}{2},m+2j+2} \varphi_{q_a+\nicefrac{1}{2},q_b+\nicefrac{1}{2}}^{j+\nicefrac{1}{2},m} \\
    \mathfrak{K}^1 \varphi_{q_a,q_b}^{j,m} &= -iC_{-q_a+\nicefrac{1}{2},q_b+\nicefrac{1}{2}}^{j+\nicefrac{1}{2},m+2j+2} \varphi_{q_a-\nicefrac{1}{2},q_b+\nicefrac{1}{2}}^{j+\nicefrac{1}{2},m} -i C_{q_a+\nicefrac{1}{2},-q_b+\nicefrac{1}{2}}^{j+\nicefrac{1}{2},m+2j+2} \varphi_{q_a+\nicefrac{1}{2},q_b-\nicefrac{1}{2}}^{j+\nicefrac{1}{2},m} \numberthis\\
    &\quad +i C_{q_a,-q_b}^{j,m+1} \varphi_{q_a-\nicefrac{1}{2},q_b+\nicefrac{1}{2}}^{j-\nicefrac{1}{2},m+1} +i C_{-q_a,q_b}^{j,m+1} \varphi_{q_a+\nicefrac{1}{2},q_b-\nicefrac{1}{2}}^{j-\nicefrac{1}{2},m+1} \\
    \mathfrak{K}^2 \varphi_{q_a,q_b}^{j,m} &= -C_{-q_a+\nicefrac{1}{2},q_b+\nicefrac{1}{2}}^{j+\nicefrac{1}{2},m+2j+2} \varphi_{q_a-\nicefrac{1}{2},q_b+\nicefrac{1}{2}}^{j+\nicefrac{1}{2},m} + C_{q_a+\nicefrac{1}{2},-q_b+\nicefrac{1}{2}}^{j+\nicefrac{1}{2},m+2j+2} \varphi_{q_a+\nicefrac{1}{2},q_b-\nicefrac{1}{2}}^{j+\nicefrac{1}{2},m} \numberthis \label{eq:action_k2}\\
    &\quad + C_{q_a,-q_b}^{j,m+1} \varphi_{q_a-\nicefrac{1}{2},q_b+\nicefrac{1}{2}}^{j-\nicefrac{1}{2},m+1} - C_{-q_a,q_b}^{j,m+1} \varphi_{q_a+\nicefrac{1}{2},q_b-\nicefrac{1}{2}}^{j-\nicefrac{1}{2},m+1} \\
    \mathfrak{K}^3 \varphi_{q_a,q_b}^{j,m} &= -iC_{q_a+\nicefrac{1}{2},q_b+\nicefrac{1}{2}}^{j+\nicefrac{1}{2},m+2j+2} \varphi_{q_a+\nicefrac{1}{2},q_b+\nicefrac{1}{2}}^{j+\nicefrac{1}{2},m} +i C_{-q_a+\nicefrac{1}{2},-q_b+\nicefrac{1}{2}}^{j+\nicefrac{1}{2},m+2j+2} \varphi_{q_a-\nicefrac{1}{2},q_b-\nicefrac{1}{2}}^{j+\nicefrac{1}{2},m} \numberthis\\
    &\quad -i C_{-q_a,-q_b}^{j,m+1} \varphi_{q_a+\nicefrac{1}{2},q_b+\nicefrac{1}{2}}^{j-\nicefrac{1}{2},m+1} +i C_{q_a,q_b}^{j,m+1} \varphi_{q_a-\nicefrac{1}{2},q_b-\nicefrac{1}{2}}^{j-\nicefrac{1}{2},m+1}
\end{align*}
\end{subequations}
with coefficients
\begin{equation}
    C_{q_a,q_b}^{j,m} = \frac{\sqrt{(j+q_a)(j+q_b)m(\mathbb{\Delta}+m-2)}}{\sqrt{2j(2j+1)}}\,.
\end{equation}

Finally, the non-vanishing components of the tensor $\mathfrak{M}^{\mu\nu}$ generating Lorentz transformations act as
\begin{align}
    \mathfrak{M}^{03} \varphi_{q_a,q_b}^{j,m} &= -i(q_a+q_b)\varphi_{q_a,q_b}^{j,m} & \mathfrak{M}^{12} \varphi_{q_a,q_b}^{j,m} &= -i(q_a-q_b)\varphi_{q_a,q_b}^{j,m}
\end{align}
\vspace{-0.5cm}
\begin{subequations}
\begin{align*}
\label{eq:M_on_Phi}
    \mathfrak{M}^{01} \varphi_{q_a,q_b}^{j,m} &= \frac{-i}{2}\Big(\sqrt{(j+q_a+1)(j-q_a)}\varphi_{q_a+1,q_b}^{j,m} + \sqrt{(j+q_a)(j-q_a+1)}\varphi_{q_a-1,q_b}^{j,m} \numberthis\\
    &\qquad+ \sqrt{(j+q_b+1)(j-q_b)}\varphi_{q_a,q_b+1}^{j,m} + \sqrt{(j+q_b)(j-q_b+1)}\varphi_{q_a,q_b-1}^{j,m} \Big) \\
    \mathfrak{M}^{02} \varphi_{q_a,q_b}^{j,m} &= \frac{1}{2}\Big(\sqrt{(j+q_a+1)(j-q_a)}\varphi_{q_a+1,q_b}^{j,m} - \sqrt{(j+q_a)(j-q_a+1)}\varphi_{q_a-1,q_b}^{j,m} \numberthis\\
    &\qquad- \sqrt{(j+q_b+1)(j-q_b)}\varphi_{q_a,q_b+1}^{j,m} + \sqrt{(j+q_b)(j-q_b+1)}\varphi_{q_a,q_b-1}^{j,m} \Big) \\
    \mathfrak{M}^{13} \varphi_{q_a,q_b}^{j,m} &= \frac{-1}{2}\Big(\sqrt{(j+q_a+1)(j-q_a)}\varphi_{q_a+1,q_b}^{j,m} - \sqrt{(j+q_a)(j-q_a+1)}\varphi_{q_a-1,q_b}^{j,m} \numberthis\\
    &\qquad+ \sqrt{(j+q_b+1)(j-q_b)}\varphi_{q_a,q_b+1}^{j,m} - \sqrt{(j+q_b)(j-q_b+1)}\varphi_{q_a,q_b-1}^{j,m} \Big) \\
    \mathfrak{M}^{23} \varphi_{q_a,q_b}^{j,m} &= \frac{-i}{2}\Big(\sqrt{(j+q_a+1)(j-q_a)}\varphi_{q_a+1,q_b}^{j,m} + \sqrt{(j+q_a)(j-q_a+1)}\varphi_{q_a-1,q_b}^{j,m} \numberthis\\
    &\qquad- \sqrt{(j+q_b+1)(j-q_b)}\varphi_{q_a,q_b+1}^{j,m} - \sqrt{(j+q_b)(j-q_b+1)}\varphi_{q_a,q_b-1}^{j,m} \Big) 
\end{align*}
\end{subequations}

\subsection{Computation of Unitarity Bounds}
Using the action of the generators on the basis, we can straightforwardly compute the unitarity bounds through
\begin{align}
    \| P_{\mu_N} \cdots P_{\mu_2} P_{\mu_1} \ket{\mathbb{\Delta}}\| &= \bra{\mathbb{\Delta}} \left( P_{\mu_N} \cdots P_{\mu_2} P_{\mu_1}\right)^\dagger  P_{\mu_N} \cdots P_{\mu_2} P_{\mu_1} \ket{\mathbb{\Delta}}\notag\\  
    &= \int_{\mathbb{D}_4} [\d U]_\Deltabb \bra{\mathbb{\Delta}} K_{\mu_1} \cdots K_{\mu_N} \ket*{U^\dagger} \bra{U} P_{\mu_N} \cdots P_{\mu_2} P_{\mu_1} \ket{\mathbb{\Delta}} \notag\\
    &=  \int_{\mathbb{D}_4} [\d U]_\Deltabb\, \overline{\bra{U} P_{\mu_N} \cdots P_{\mu_2} P_{\mu_1} \ket{\mathbb{\Delta}}} \bra{U} P_{\mu_N} \cdots P_{\mu_2} P_{\mu_1} \ket{\mathbb{\Delta}}\notag\\
      &= \int_{\mathbb{D}_4} [\d U]_\Deltabb\, \overline{  \mathfrak{K}_{\mu_N}^{(w)} \cdots \mathfrak{K}_{\mu_2}^{(w)} \mathfrak{K}_{\mu_1}^{(w)} 1} \,\,\mathfrak{K}_{\mu_N}^{(w)} \cdots \mathfrak{K}_{\mu_2}^{(w)} \mathfrak{K}_{\mu_1}^{(w)} 1\notag\\
      &=
      \left(\mathfrak{K}_{\mu_N}^{(w)} \cdots \mathfrak{K}_{\mu_2}^{(w)} \mathfrak{K}_{\mu_1}^{(w)} 1\,,\mathfrak{K}_{\mu_N}^{(w)} \cdots \mathfrak{K}_{\mu_2}^{(w)} \mathfrak{K}_{\mu_1}^{(w)} 1\right) \notag\\ &\geq 0\,,
\end{align}
where we used equation \eqref{eq:draw_out_generator_4d} together with equation \eqref{eq:adjoint_relations_4d}. Since $\Deltabb \in \mathbb{R}$ we have $\mathfrak{K}_{\mu_i}^{(\Bar{w})} = \overline{\mathfrak{K}_{\mu_i}^{(w)}}$. Furthermore, the semi-definiteness in the last step follows from the action of the SCTs on the normalised basis in equation \eqref{eq:K_on_Phi} and $\varphi_{0,0}^{0,0} = 1$.

\section{Correspondence with Literature}
In this appendix we first provide computational details to the induction of the $n$-point comb block, which we started in equation \eqref{eq:induction_npt_comb}. Next, a brief review of the conformal four-point block in the case of an internal primary with scaling dimension $\Deltabb$ and spin $j$ is presented. We demonstrate that the result derived through the oscillator formalism in section \ref{sec:4ptblock_4d} aligns with the one in the literature for the scalar case.

\subsection{\texorpdfstring{The $n$-point Comb Block in Two Dimensions}{The n-point Comb Block in Two Dimensions}}
\label{app:comb_npt}

To inductively evaluate \eqref{eq:induction_npt_comb}, we plug in the result for the $(n-1)$-point comb block according to equation \eqref{eq:comb_npt_block} and obtain for the $n$-point block
\begin{align}
    & G^{h_1,\dots,h_n}_{\mathbb{h}_1,\dots,\mathbb{h}_{n-3}}(z_1,\dots,z_n)  \\
    =& \int_\mathbb{D}[\d^2u] \left(\frac{z_{23}}{z_{12}z_{13}}\right)^{h_1} z_{n-3,n-2}^{\hbb_{n-3}-h_{n-2}}(z_{n-3}-\bar{u})^{-\hbb_{n-3}+h_{n-2}}(z_{n-2}-\bar{u})^{-\hbb_{n-3}-h_{n-2}}\prod\limits_{i=1}^{n-5}\xi_i^{\mathbb{h}_i}\left(\xi_{\bar{u}}\right)^{\mathbb{h}_{n-4}}\nonumber\\
    & \cdot\prod\limits_{i=1}^{n-4} \left(\frac{z_{i,i+2}}{z_{i,i+1}z_{i+1,i+2}}\right)^{h_{i+1}}\!\!\!
    \psi(z_{n-1},z_n;u)  \nonumber F_K\bigg[
    \begin{matrix}
    h_{112}, \hbb_{123},\dots,\hbb_{n-4,n-3,n-2}\\ 2\mathbb{h}_1,\dots,2\mathbb{h}_{n-4}
    \end{matrix}; \xi_1,\dots, \xi_{n-5},\xi_{\bar{u}}\bigg]
\end{align}
with $\hbb_{ijk}\equiv\hbb_i+\hbb_j-h_k$, $h_{ijk}\equiv\hbb_i+h_j-h_k$ and $\xi_{\bar{u}}\equiv\frac{z_{n-4,n-3}(z_{n-2}-\bar{u}_{n-3})}{z_{n-4,n-2}(z_{n-3}-\bar{u}_{n-3})}$. Note that $\xi_{\bar{u}}$ is just the $(n-4)$th cross-ratio of the $(n-1)$-point block, where the last point $z_{n-1}$ was replaced with an oscillator variable $\bar{u}$. Next, we set $z_{n-1} = 0$ and take the limit $z_{n-2}\to\infty$. 
In this choice, $\xi_{\bar{u}}= \frac{z_{n-4,n-3}}{z_{n-3}-\bar{u}_{n-3}}$ and the last three cross-ratios read
\begin{equation}
    \xi_{n-5} = \frac{z_{n-5,n-4}}{z_{n-5,n-3}}, \quad \xi_{n-4}= -\frac{z_{n-4,n-3}}{z_{n-3}}, \quad \xi_{n-3} = \frac{z_n}{z_{n-3}}, \label{eq:SL2R_npoint_cross_ratios}
\end{equation}
whereas the other cross-ratios remain unchanged. Furthermore, we expand the remaining factors that depend on oscillator variables into monomials
\begin{align}
    & G^{h_1,\dots,h_n}_{\mathbb{h}_1,\dots,\mathbb{h}_{n-3}}(z_1,\dots,z_{n-2}\to \infty,0,z_n)  \\
    =& \left(\frac{z_{23}}{z_{12}z_{13}}\right)^{h_1} \prod\limits_{i=1}^{n-4} \left(\frac{z_{i,i+2}}{z_{i,i+1}z_{i+1,i+2}}\right)^{h_{i+1}}\!\! \left(\frac{-z_{n-3}}{z_{n-2}^2}\right)^{h_{n-2}} (-z_n)^{-h_{n-1}-h_n} \prod\limits_{i=1}^{n-3}\xi_i^{\mathbb{h}_i} \nonumber\\
    & \cdot\!\!\sum\limits_{k_1,\dots,k_{n-2}=0}^\infty \!\!\!\!\!\!\!\!(-1)^{k_{n-3}+k_{n-2}}\frac{\xi_1^{k_1}}{k_1!}\dots \frac{\xi_{n-4}^{k_{n-4}}}{k_{n-4}!} \binom{-\hbb_{n-3,n-4,n-2}-k_{n-4}}{k_{n-3}} \binom{-h_{n-3,n,n-1}}{k_{n-2}} (h_{112})_{k_1}\nonumber\\
    & \cdot\frac{ (\hbb_{123})_{k_1+k_2} \dots (\hbb_{n-5,n-4,n-3})_{k_{n-5}+k_{n-4}}  }{(2\mathbb{h}_1)_{k_1}\dots(2\mathbb{h}_{n-4})_{k_{n-4}}}   (\hbb_{n-4,n-3,n-2})_{k_{n-4}} z_{n-3}^{-k_{n-3}} z_n^{k_{n-2}} \cdot\! \int_\mathbb{D}[\d^2u] \bar{u}^{k_{n-3}}u^{k_{n-2}}\,. \notag
\end{align}
We perform the integral making use of orthogonality to terminate the sum over $k_{n-2}$ and obtain the expression
\begin{align}
    &\left(\frac{z_{23}}{z_{12}z_{13}}\right)^{h_1} \prod\limits_{i=1}^{n-4} \left(\frac{z_{i,i+2}}{z_{i,i+1}z_{i+1,i+2}}\right)^{h_{i+1}}\!\! \left(\frac{-z_{n-3}}{z_{n-2}^2}\right)^{h_{n-2}} (-z_n)^{-h_{n-1}-h_n} \prod\limits_{i=1}^{n-3}\xi_i^{\mathbb{h}_i} \nonumber\\
    & \cdot\sum\limits_{k_1,\dots,k_{n-3}=0}^\infty \frac{\xi_1^{k_1}}{k_1!}\dots \frac{\xi_{n-4}^{k_{n-4}}}{k_{n-4}!} \binom{-\hbb_{n-3,n-4,n-2}-k_{n-4}}{k_{n-3}}\! \binom{-h_{n-3,n,n-1}}{k_{n-3}} (h_{112})_{k_1}(\hbb_{n-4,n-3,n-2})_{k_{n-4}}\nonumber\\
    & \cdot\frac{ (\hbb_{123})_{k_1+k_2} \dots (\hbb_{n-5,n-4,n-3})_{k_{n-5}+k_{n-4}}  }{(2\mathbb{h}_1)_{k_1}\dots(2\mathbb{h}_{n-3})_{k_{n-3}}}    \xi_{n-3}^{k_{n-3}}  \cdot k_{n-3}!
\end{align}
In the last step, we rearranged powers of points and their differences in a convenient way such that we obtain the cross-ratios from \eqref{eq:SL2R_npoint_cross_ratios}. 

Writing the binomial coefficients in terms of Pochhammer symbols and rearranging gives us the final result
\begin{align}
    & G^{h_1,\dots,h_n}_{\mathbb{h}_1,\dots,\mathbb{h}_{n-3}}(z_1,\dots,z_{n-2}\to \infty,0,z_n) \nonumber \\
    =& \left(\frac{z_{23}}{z_{12}z_{13}}\right)^{h_1} \prod\limits_{i=1}^{n-4} \left(\frac{z_{i,i+2}}{z_{i,i+1}z_{i+1,i+2}}\right)^{h_{i+1}}\!\! \left(\frac{-z_{n-3}}{z_{n-2}^2}\right)^{h_{n-2}} (-z_n)^{-h_{n-1}-h_n}  \prod\limits_{i=1}^{n-3}\xi_i^{\mathbb{h}_i} \nonumber\\
    & \cdot\sum\limits_{k_1,\dots,k_{n-3}=0}^\infty \frac{\xi_1^{k_1}}{k_1!}\dots \frac{\xi_{n-3}^{k_{n-3}}}{k_{n-3}!} \cdot\frac{ (h_{112})_{k_1}(\hbb_{123})_{k_1+k_2} \dots  (\hbb_{n-4,n-3,n-2})_{k_{n-4}+k_{n-3}} (h_{n-3,n,n-1})_{k_{n-3}}}{(2\mathbb{h}_1)_{k_1}\dots(2\mathbb{h}_{n-3})_{k_{n-3}}} \notag\\
    =& \mathcal{L}^{h_1,\dots,h_n}(z_1,\dots,z_{n-2}\to \infty,0,z_n) \,g^{h_1,\dots,h_n}_{\mathbb{h}_1,\dots,\mathbb{h}_{n-3}}(z_1,\dots,z_{n-2}\to \infty,0,z_n).
\end{align}
This perfectly matches the result from \cite{Rosenhaus:2018zqn}.

\subsection{The Scalar Four-Point Block in Four Dimensions}
\label{app:correspondence_4p}
The four-point block with spinning inner weight in four dimensions is computed in \cite{Dolan:2000ut,Dolan:2003hv}. Let us quickly review the results for comparison. For four points $x_0,x_1,x_2,x_3$, there exist two independent cross-ratios $u$ and $v$,
\begin{align}
    u &= \frac{x_{12}^2x_{34}^2}{x_{13}^2 x_{24}^2}\,, & v &= \frac{x_{14}^2 x_{23}^2}{x_{13}^2 x_{24}^2}\,, \label{eq:4pt_cross_ratios}
\end{align}
where we adapted the usual notation $x_{ij} = \abs*{x_i-x_j}$. Moreover, we can reparameterise the cross-ratios in terms of new variables $(z,\Bar{z})$ via $u = z \Bar{z}$ and $v=(1-z)(1-\Bar{z})$. Expressed through these, the four-point block is given by 
\begin{align}
    G^{\Delta_1,\dots,\Delta_4}_{(\Deltabb,j)} &= \frac{1}{z-\Bar{z}}\left( z^{\lambda_1+1}\Bar{z}^{\lambda_2}{}_2F_1\left( \begin{matrix}
         \lambda_1 +a, \lambda_1 + b \\ 2\lambda_1 
     \end{matrix}; z \right)  {}_2F_1  \left( \begin{matrix}
         \lambda_2 +a-1, \lambda_2 + b-1 \\ 2\lambda_2  - 2
     \end{matrix}; \Bar{z} \right)\right.\notag\\
     &\hspace{2cm}  - (z \leftrightarrow \Bar{z})\
     \Biggl)\label{eq:DolanOsborn_4ptBlock}
\end{align}
with parameters 
\begin{align}
\begin{split}
    \lambda_1 &= \frac{1}{2}(\Deltabb+j)\,, \qquad a = \frac{1}{2}(\Delta_2-\Delta_1)\,,  \\
     \lambda_2 &= \frac{1}{2} (\Deltabb-j)\,, \qquad b = \frac{1}{2}(\Delta_3-\Delta_4)\,,   
\end{split}
\end{align}
where $(\Deltabb,j_1=j, j_2=j)$ are scaling dimension and spin of the internal primary. Note that the second summand in equation \eqref{eq:DolanOsborn_4ptBlock} is necessary to make the block manifestly symmetric in the two variables. In the scalar case $j =0$, one has $\lambda_1=\lambda_2=\frac{\Deltabb}{2}$ and thus the first part of the conformal block reads 
\begin{equation}
    \frac{(z \Bar{z})^{\frac{\Deltabb}{2}}}{z-\Bar{z}}z\,\, {}_2F_1 \left( \begin{matrix}
        \frac{1}{2}\Deltabb+a, \frac{1}{2}\Deltabb+b \\ \Deltabb
    \end{matrix}; z \right)\,{}_2F_1 \left( \begin{matrix}
        \frac{1}{2}\Deltabb+a-1, \frac{1}{2}\Deltabb+b-1 \\ \Deltabb -2
    \end{matrix}; \Bar{z} \right)\,. \label{eq:DolanOsborn_4ptBlock_scalar}
\end{equation}

In the following, we show that our result in equation \eqref{eq:result_4pt_4d} actually matches equation \eqref{eq:DolanOsborn_4ptBlock_scalar}.
Setting $k = 2j+1$, we express the coefficients of \eqref{eq:result_4pt_4d} in terms of Pochhammer symbols as follows 
\begin{align}
\left(\mathcal{N}^{j,m}_{\beta,\varepsilon;\Deltabb}\right)^2 &= \left(\mathcal{N}^{\frac{k-1}{2},m}_{\beta}\right)^2 \left(\mathcal{N}^{\frac{k-1}{2},m}_{\varepsilon}\right)^2 \left(\mathcal{N}^{\frac{k-1}{2},m}_{\Deltabb}\right)^{-2}\notag \\
    &= \frac{(\Deltabb-1)\,k}{(\beta-1)(\varepsilon-1)}\frac{\binom{m+\beta-2}{\beta-2} \binom{m+k+\beta-2}{\beta-2} 
    \binom{m+\varepsilon-2}{\varepsilon-2}\binom{m+k+\varepsilon-2}{\varepsilon-2}}{ 
    \binom{m+\Deltabb-2}{\Deltabb-2} \binom{m+k+\Deltabb-2}{\Deltabb-2}} \\
    &= \frac{k}{m!(m+k)!} \frac{\Deltabb-1}{(\beta-1)(\varepsilon-1)} \frac{(\beta-1)_{m}(\beta-1)_{m+k} (\varepsilon-1)_{m}(\varepsilon-1)_{m+k}}{(\Deltabb-1)_{m}(\Deltabb-1)_{m+k}}\,. \nonumber
\end{align}
In the last step, we used the identity $\binom{a+m}{a} = \frac{(a+1)_m}{m!}$.
Next, we expand the first factor and, after some rearranging and using that $(a)_{m-1} = \frac{(a-1)_m}{a-1} $, arrive at
\begin{align}
\begin{split} \label{eq:app_N_manipulation2}
    \left(\mathcal{N}^{j,m}_{\beta,\varepsilon;\Deltabb}\right)^2  
    &= \left(\frac{1}{(m+k-1)!m!} - \frac{1}{(m-1)!(m+k)!}\right) \frac{\Deltabb-1}{(\beta-1)(\varepsilon-1)} \\
    &\qquad\cdot \frac{(\beta-1)_{m}(\beta-1)_{m+k} (\varepsilon-1)_{m}(\varepsilon-1)_{m+k}}{(\Deltabb-1)_{m}(\Deltabb-1)_{m+k}}\\
    &= \frac{1}{(m+k-1)!m!} \frac{(\beta)_{m+k-1}(\varepsilon)_{m+k-1}(\beta-1)_{m}(\varepsilon-1)_{m}}{(\Deltabb)_{m+k-1}(\Deltabb-1)_{m}} \\
    &\quad- \frac{1}{(m-1)!(m+k)!} \frac{(\beta)_{m-1}(\varepsilon)_{m-1}(\beta-1)_{m+k}(\varepsilon-1)_{m+k}}{(\Deltabb)_{m-1}(\Deltabb-1)_{m+k}}\,.
\end{split}
\end{align}
Inserting helpful zeros of the form $\frac{1}{(\Deltabb-2)_{m}}-\frac{1}{(\Deltabb-2)_{m}}$ in the first summand and $\frac{1}{(\Deltabb-2)_{m+k}}-\frac{1}{(\Deltabb-2)_{m+k}}$ in the second summand, \eqref{eq:app_N_manipulation2} amounts to 
\begin{align}
    &= \frac{(\beta)_{m+k-1}(\varepsilon)_{m+k-1}(\beta-1)_{m}(\varepsilon-1)_{m}}{(m+k-1)!m!(\Deltabb)_{m+k-1}}\left(\frac{1}{(\Deltabb-1)_{m}}-\frac{1}{(\Deltabb-2)_{m}}+\frac{1}{(\Deltabb-2)_{m}}\right) \\
    &\quad- \frac{(\beta)_{m-1}(\varepsilon)_{m-1}(\beta-1)_{m+k}(\varepsilon-1)_{m+k}}{(m-1)!(m+k)!(\Deltabb)_{m-1}}\left(\frac{1}{(\Deltabb-1)_{m+k}}-\frac{1}{(\Deltabb-2)_{m+k}}+\frac{1}{(\Deltabb-2)_{m+k}}\right)\,.\notag
\end{align}
Finally, one arrives at
\begin{align}
\begin{split}
\left(\mathcal{N}^{j,m}_{\beta,\varepsilon;\Deltabb}\right)^2  
    &= \frac{1}{(m+k-1)!m!} \frac{(\beta)_{m+k-1}(\varepsilon)_{m+k-1}(\beta-1)_{m}(\varepsilon-1)_{m}}{(\Deltabb)_{m+k-1}(\Deltabb-2)_{m}} \\
    &\quad- \frac{1}{(m-1)!(m+k)!} \frac{(\beta)_{m-1}(\varepsilon)_{m-1}(\beta-1)_{m+k}(\varepsilon-1)_{m+k}}{(\Deltabb)_{m-1}(\Deltabb-2)_{m+k}}\\
    &= \mathcal{C}_{\beta,\varepsilon;\Deltabb}^{m+k-1;m}-\mathcal{C}_{\beta,\varepsilon;\Deltabb}^{m-1;m+k}\,,
\end{split}
\end{align}
where we introduced the coefficients\footnote{Note that these coefficients become ill-defined for $a,b=-1$, however, we can simply set those coefficients to zero. In particular, with this convention equation \eqref{eq:app_N_manipulation2} holds even for $m=0$.} 
\begin{equation}\label{eq:coefficients_C}
    \mathcal{C}_{\beta,\varepsilon;\Deltabb}^{a;b} = \frac{1}{a!}\frac{(\beta)_a(\varepsilon)_a}{(\Deltabb)_a}\, \frac{1}{b!}\frac{(\beta-1)_{b}(\varepsilon-1)_{b}}{(\Deltabb-2)_{b}}\,.
\end{equation}
Plugging this result into the sum \eqref{eq:result_4pt_4d} and performing elementary index shifts, we find
\begin{align}
    &\frac{(z\bar{z})^{\frac{\Deltabb}{2}}}{z-\bar{z}} \sum\limits_{m=0}^\infty \sum\limits_{j\in\mathbb{N}/2}  \left(\mathcal{N}^{j,m}_{\beta,\varepsilon;\Deltabb}\right)^2 z^m \bar{z}^m (z^{2j+1} - \bar{z}^{2j+1}) \nonumber \\
    &\qquad= \frac{(z\bar{z})^{\frac{\Deltabb}{2}}}{z-\bar{z}} \sum\limits_{m=0}^\infty \sum\limits_{k=1}^\infty \left(\mathcal{C}_{\beta,\varepsilon;\Deltabb}^{m+k-1;m}-\mathcal{C}_{\beta,\varepsilon;\Deltabb}^{m-1;m+k}\right) z^m \bar{z}^m (z^{k} - \bar{z}^{k})\\
    &\qquad= \frac{(z\bar{z})^{\frac{\Deltabb}{2}}}{z-\bar{z}} \left( \sum\limits_{b=0}^\infty \sum\limits_{a=b}^\infty \mathcal{C}_{\beta,\varepsilon;\Deltabb}^{a;b} (z^{a+1}\bar{z}^b - z^b\bar{z}^{a+1})  - \sum\limits_{a=0}^\infty \sum\limits_{b=a+1}^\infty \mathcal{C}_{\beta,\varepsilon;\Deltabb}^{a;b} (z^{b}\bar{z}^{a+1} - z^{a+1}\bar{z}^{b}) \right) \nonumber\,.
\end{align}
By rearranging the summations, we can simplify the expression to
\begin{align}
 &\frac{(z\bar{z})^{\frac{\Deltabb}{2}}}{z-\bar{z}} \sum\limits_{m=0}^\infty \sum\limits_{j\in\mathbb{N}/2}  \left(\mathcal{N}^{j,m}_{\beta,\varepsilon;\Deltabb}\right)^2 z^m \bar{z}^m (z^{2j+1} - \bar{z}^{2j+1}) \nonumber \\
    &\qquad= \frac{(z\bar{z})^{\frac{\Deltabb}{2}}}{z-\bar{z}} \sum\limits_{a=0}^\infty \sum\limits_{b=0}^\infty \mathcal{C}_{\beta,\varepsilon;\Deltabb}^{a;b} (z^{a+1}\bar{z}^b - z^{b}\bar{z}^{a+1}) \\
    &\qquad= \frac{(z\bar{z})^{\frac{\Deltabb}{2}}}{z-\bar{z}} \left( z\, {}_2F_1 \left( \begin{matrix}
        \beta, \varepsilon \\ \Deltabb
    \end{matrix}; z \right) \,{}_2F_1 \left( \begin{matrix}
        \beta - 1, \varepsilon - 1 \\ \Deltabb -2
    \end{matrix}; \bar{z} \right) 
    -(z\leftrightarrow \Bar{z}) \right)\,,\nonumber
\end{align}
which concludes the comparison.

\end{appendices}

\bibliography{main}
\bibliographystyle{JHEP}

\end{document}